\documentclass[preprint2]{aastex6}

\usepackage{ulem}
\usepackage{xcolor}
\usepackage{amsmath}
\usepackage{cases}
\usepackage[T1]{fontenc}
\usepackage{subfigure}
\newcommand{\Tbz}{T_{\rm{b,z}}}
\newcommand{\Lbz}{L_{\rm{b,z}}}
\newcommand{\Epi}{E_{\rm{p,i}}}
\newcommand{\Eiso}{E_{\gamma,\rm{iso}}}
\newcommand{\ext}{\sigma_{\rm{int}}}
\newcommand{\dl}{d_{\rm{L}}}
\newcommand{\om}{\Omega_{\rm m}}
\newcommand{\oL}{\Omega_{\rm \Lambda}}

\newcommand{\LT}{\emph{L-T}}
\newcommand{\LTE}{\emph{L-T-E}}
\newcommand{\LTEpi}{\emph{L-T-E$_{\rm{p,i}}$}}
\newcommand{\LTEiso}{\emph{L-T-E$_{\rm{\gamma,iso}}$}}

\begin{document}
\title{Radio Plateaus in Gamma-Ray Burst Afterglows and Their Application in Cosmology}
\author{Xiao Tian\altaffilmark{1}, Jia-Lun Li\altaffilmark{1}, Shuang-Xi Yi\altaffilmark{1},Yu-Peng Yang\altaffilmark{1}, Jian-Ping Hu\altaffilmark{2}, Yan-Kun Qu\altaffilmark{1} and Fa-Yin Wang\altaffilmark{2}}
\altaffiltext{1}{School of Physics and Physical Engineering, Qufu Normal University, Qufu 273165, China; yisx2015@qfnu.edu.cn}
\altaffiltext{2}{School of Astronomy and Space Science, Nanjing University, Nanjing 210023, China; fayinwang@nju.edu.cn}

\begin{abstract}
The plateau phase in the radio afterglows has been observed in very few gamma-ray bursts (GRBs), and 27 radio light curves with plateau phase were acquired from the published literature in this article.
We obtain the related parameters of the radio plateau, such as temporal indexes during the plateau phase ($\alpha_1$ and $\alpha_2$), break time ($\Tbz$) and the corresponding radio flux ($F_{\rm b}$). The two parameter Dainotti relation between the break time of the plateau and the corresponding break luminosity ($\Lbz$) in radio band is $\Lbz \propto \Tbz^{-1.20\pm0.24}$. Including the isotropic energy $\Eiso$ and the peak energy $\Epi$, the three parameter correlations for the radio plateaus are written as $\Lbz \propto \Tbz^{-1.01 \pm 0.24} \Eiso^{0.18 \pm 0.09}$ and $\Lbz \propto \Tbz^{-1.18 \pm 0.27} \Epi^{0.05 \pm 0.28}$, respectively. The correlations are less consistent with that of X-ray and optical plateaus, implying that radio plateaus may have a different physical mechanism. The typical frequencies crossing the observational band may be a reasonable hypothesis that causes the breaks of the radio afterglows. We calibrate GRBs empirical luminosity correlations as standard candle for constraining cosmological parameters, and find that our samples can constrain the flat $\Lambda$CDM model well, while are not sensitive to non-flat ${\Lambda}$CDM model. By combining GRBs with other probes, such as SN and CMB, the constraints on cosmological parameters are $\om = 0.297\pm0.006$ for the flat ${\Lambda}$CDM model and $\om = 0.283\pm0.008$, $\oL = 0.711\pm0.006$ for the non-flat ${\Lambda}$CDM model, respectively.
\end{abstract}

\section{Introduction}
\label{Sec:A: Introduction}
Gamma-ray bursts are deemed extremely high-energy events in the universe, lasting from ten milliseconds to several hours, usually with the isotropic energy emitted in the prompt emission vary from $10^{\rm 48}$ to $10^{\rm 55}$ erg \citep{2015PhR...561....1K, 2011ApJ...732...29C}. The most popular model proposed to explain the origin of GRBs is the standard fireball model\citep{2004RvMP...76.1143P, 2006RPPh...69.2259M, 2007ChJAA...7....1Z, 2015PhR...561....1K}. The prompt $\gamma$-ray emission can be explained by the internal shocks caused by the interaction of ejecta in the fireball. While when the fireball ejecta is decelerated by the ambient medium, the interaction between the shocks produces the multi-band afterglows, e.g., X-ray, optical and radio \citep{1993ApJ...405..278M, 1997ApJ...476..232M, 1998ApJ...497L..17S,2000MNRAS.316..943H,2013ApJ...776..120Y,2018ApJ...859..160W,2020ApJ...895...94Y,2021ApJ...908..242D}.

GRBs can be simply classified into two categories based on the observed prompt emission duration ($T_{90}$) and spectral hardness\citep{2013ApJ...763...15Q, 2020ApJ...893...77W}. Most long GRBs ($T_{\rm 90} > 2s$) are believed to birth from the collapse of massive stars \citep{1993ApJ...405..273W, 1999ApJ...524..262M}, while some long GRBs could be related to core-collapse supernovas \citep{1998Natur.395..670G, 2003ApJ...591L..17S, 2006Natur.442.1008C, 2013ApJ...776...98X}. The progenitors of short GRBs ($T_{\rm 90} < 2s$) are thought to be from the merger of binary compact stars, i.e., NS-NS \citep{1986ApJ...308L..43P, 1989Natur.340..126E} or NS-BH \citep{2002ApJ...571..779P, 1991AcA....41..257P}. Both models support a stellar-size, hyper-accreting BH or a rapidly spinning, strongly magnetized NS as the central engine of GRBs (\citealt{1994MNRAS.270..480T}; \citealt{1998A&A...333L..87D}a; \citealt{1999ApJ...518..356P, 2003MNRAS.345.1077R, 2013ApJ...765..125L}; \citealt{2017ApJ...850...30L, 2018ApJ...852...20L, 2021MNRAS.507.1047Y}). It is noteworthy that a large portion of X-ray afterglows present a plateau emission at early stages, which is caused by the continuous energy injection from a central engine(\citealt{1998PhRvL..81.4301D}b; \citealt{2005MNRAS.364L..42F, 2013MNRAS.430.1061R, 2013NatPh...9..465W, 2015ApJ...813...92R, 2016ApJS..224...20Y, 2017ApJ...844...79Y, 2022ApJ...924...69Y}). The fallback of matter onto a newly formed black hole or the spin-down luminosity of a newborn magnetar are two possible models  interpreting the energy injection. \cite{2014MNRAS.443.1779R} interpreted the Dainotti relation in X-rays by simulating the plausible values of magnetic field and spin period, and their work  supports the magnetar model. \cite{2011A&A...526A.121D} and \cite{2018ApJ...869..155S} also explained the X-ray plateau with magnetar model, providing a reliable physical basis.

Similar plateau feature appears in the radio afterglow. However, the sample size of radio afterglow with plateau phases are much smaller than that of X-ray and optical, which should be related to the low detection rate of radio afterglows. \cite{2012ApJ...746..156C} showed that only about 31$\%$ of long GRBs have detectable radio afterglows, compared to $\sim$ 95$\%$ and $\sim$ 70$\%$ of X-ray and optical afterglows, respectively. \cite{2023MNRAS.520.5764C} analyzed 211 GRBs with redshift measurements and divided them into two distinct classes: radio-bright GRBs (123, with radio afterglow emission) and radio-dark GRBs (88, without radio afterglow emission), showing a greatly improved detection rate of the radio afterglows.

It is well konwn that many astronomical observations, such as Type Ia supernovae (SNe Ia), the cosmic microwave background (CMB) and the baryon acoustic oscillations (BAO) can be used to constrain cosmological parameters (\citealt{1993ApJ...413L.105P, 1998AJ....116.1009R, 1999ApJ...517..565P,2003ApJS..148..175S,2005ApJ...633..560E, 2005MNRAS.362..505C, 2014MNRAS.441...24A}; \citealt{2023ApJ...951...63D}b). A called Lambda cold dark matter ($\Lambda$CDM) model have been presented by these studies. The mechanism of SNe Ia limits its observed redshift to an upper limit of about $z\sim 2.26$ \citep{2018ApJ...859..101S}. The informations about the high redshift of the early universe around $z\sim 1089$ can be provided by the anisotropy of the CMB. Therefore, there is a redshift blankness between the SNe Ia and CMB. Due to the cosmological origin and wide redshift distribution (up to $z\sim$ 9.40) of GRBs, they have the potential to bridge the blankness between SNe Ia and CMB. More interestingly, compared with SNe Ia observations that suffer extinction from the interstellar medium (ISM), gamma-ray photons are much less affected as they travel towards us \citep{2015NewAR..67....1W}. Therefore, GRBs have great advantages as some kinds of standard candles that constrains cosmological parameters \citep{2004ApJ...612L.101D, 2008ApJ...685..354L,2010MNRAS.408.1181C, 2016MNRAS.455.2131L, 2019MNRAS.486L..46A,2023A&A...673A..20X,2023ApJ...953...58L} and cosmic star formation rate \citep{2015ApJS..218...13Y,2018ApJ...852....1Z}. However, the process is not so easy because of the complex classifications and physical mechanisms of GRBs.

All these studies are based on empirical luminosity correlations that can be used to standardize GRBs (\citealt{2002A&A...390...81A, 2004ApJ...609..935Y, 2009MNRAS.400..775C}; \citealt{2013MNRAS.436...82D}a; \citealt{2014ApJ...783..126P, 2023MNRAS.521.3909B}; \citealt{2022MNRAS.514.1828D}a; \citealt{2022PASJ...74.1095D}b; \citealt{2023MNRAS.518.2201D}a; \citealt{2022MNRAS.512..439C}a; \citealt{2023PhRvD.107j3521C}). \cite{2010ApJ...722L.215D} mainly analyzed 77 X-ray afterglows with plateau phase and confirmed an antic-correlation that can be expressed as $\Lbz \propto \Tbz^{-1.06\pm0.27}$. This two-dimensional correlation, named 2D Dainotti relation, connects the plateau break time ($\Tbz$) and the relevant X-ray luminosity at that moment ($\Lbz$, measured in the rest frame), and was first established by \cite{2008MNRAS.391L..79D}. Interestingly, there is also an antic-correlation between the end time of the optical plateau in the rest-frame $T_{opt}$ and the corresponding luminosity $L_{opt}$ of the optical sample \citep{2012ApJ...758...27L}. The luminosity-time correlation for optical plateaus of 102 GRBs was studied by \cite{2020ApJ...905L..26D}a, $L_{opt} \propto T_{opt}^{*-1.02\pm0.16}$. The slope of this correlation is similar to the 2D Dainotti relation for X-rays and can also be explained by the magnetar model \citep{2014MNRAS.443.1779R}. \cite{2022ApJ...925...15L} reported a similar correlation for radio plateaus. Including a new parameter, the peak luminosity in the prompt emission $L_{peak}$, a three-parameter correlation was discovered ($L_{X}$-$T_{X}$-$L_{peak}$, named 3D correlation)(\citealt{2016ApJ...825L..20D, 2017A&A...600A..98D}; \citealt{2020ApJ...904...97D}b), which is an extension of the Dainotti relation. \cite{2022ApJS..261...25D}c extended the optical plateau sample and showed the existence of the three-parameter optical correlation of $L_{opt}$-$T_{opt}$-$L_{peak}$. \cite{2012A&A...538A.134X} also made an attempt to add the isotropic energy $\Eiso$ into the two-dimensional dependent $\Lbz$-$\Tbz$ correlation, and found a much tighter three-parameter correlation ($\LTE$) than the previous. \cite{2018ApJ...863...50S} compiled a group of optical plateau samples and found that $\Lbz \propto \Tbz^{\rm -0.9}\Eiso^{\rm 0.4}$ and $\Lbz \propto \Tbz^{\rm -0.9}\Epi^{\rm 0.5}$. \cite{2019ApJS..245....1T} analyzed the plateau phase in X-ray samples, and confirmed the best-fit relation $\Lbz \propto \Tbz^{\rm -1.01}\Eiso^{\rm 0.84}$. The tight correlations are expected to provide a better approach to constraint cosmological parameters.

Recently, \cite{2022ApJ...924...97W} used the total 31 long GRBs with plateau phase caused by the same physical process to explore how to limit cosmological parameters. They studied the $\LT$ correlation of X-ray plateaus, and found the fitting result is $\om= 0.34\pm0.05$(1$\sigma$) for the flat ${\Lambda}$CDM model, and $\om = 0.32^{+0.05}_{-0.10}$, $\oL = 1.10^{+0.12}_{-0.31}$(1$\sigma$) for the non-flat ${\Lambda}$CDM model. \cite{2021ApJ...920..135X} limited cosmological parameters using the $\LTEpi$ correlation of GRBs with X-ray plateau. Combining the observations of SN, BAO and CMB, the constraints on the parameters are $\om = 0.291\pm0.001$ for the flat ${\Lambda}$CDM model, and $\om = 0.289\pm0.001$, $\oL = 0.710\pm0.001$ for the non-flat ${\Lambda}$CDM model. In this work, we will extend the previous works by attempting to constrain cosmological parameters using GRBs with radio plateau.

This paper is organized as follows. We introduce the selection criteria of radio afterglows with plateau phase in Section 2. In Section 3, we analyze various correlations of GRBs. The constraints on cosmological parameters using the GRBs with radio plateau phase are shown in Section 4. We finally give the discussion and conclusion in Section 5.

\section{SAMPLE SELECTION AND LIGHT CURVES FITTING}

GRBs with plateau features propose several interesting empirical correlations that could give important prompt message to understand physical mechanisms of GRBs, and even provide a new method to constrain cosmological models. The plateau phase is a common phenomenon appeared in both X-ray and optical afterglows, which currently understood as being due to ongoing energy injection from the central engine. A fast rotating pulsar magnetar which spins down through magnetic dipole radiation as the central engine is one reasonable scenario \citep{2001ApJ...552L..35Z, 2007ApJ...670..565L, 2013MNRAS.430.1061R, 2014ApJ...785...74L}. Compared with optical and X-ray afterglows, the number of radio afterglows with plateau phase is relatively rare. The systematic collection and sorting of radio plateau samples will help us better understand the physical origin of GRBs, and study whether the radiation characteristics are consistent with the origin of X-ray and optical samples.

In this paper, we attempt to analyze the correlations of radio plateaus by an extensive search of the published literature. Well-sampled with radio plateau phases are selected from GRBs occurring between 1997 and 2022. The largest portion of our sample comes from \cite{2022ApJ...925...15L}. The sample selected should satisfy the following conditions: (1) there is a distinct flat phase in the light curve of the selected sample which can be identified as a plateau phase; (2) the selected GRBs need to have redshift measurements so that one can calculate the isotropic gamma-ray burst energy $E_{\rm \gamma,iso}$ and the intrinsic parameter $T_{\rm b,z}$ by thinking about the time dilation effect. After selecting the sample, we fit the radio plateau phases with a smooth broken power-law function (SBPL, \citealt{1999A&A...352L..26B, 2016ApJS..224...20Y}):
\begin{equation}
  F_{\rm model}(t) = F_{0} [(\frac{t}{ T_{\rm b}})^{\rm \alpha_{1}\omega}+(\frac{t}{T_{\rm b}})^{\rm \alpha_{2}\omega}]^{-\frac{1}\omega} ,
\end{equation}
where $T_{\rm b}$ is the break time, which can be obtained from the rest frame $\Tbz=T_{\rm b}/{\rm (1+z)}$. $F_{\rm b}=F_{\rm 0}\times 2^{\rm -1/{\omega}}$ is the radio flux at the break time. $\omega$ is a smoothness parameter of the light curve component with a typical value of 3.00. $\alpha_{\rm 1}$ and $\alpha_{\rm 2}$ are the temporal indexes during the plateau phase, and $\alpha_{\rm 1}$ determines the flatness of the plateau. According to the fitting results, we further remove some GRBs that are too steep to actually be plateau phases with $|\alpha_{\rm 1}| > 0.5$. The detection probability of the radio plateaus is much smaller than that of X-ray plateaus, or even lower than optical plateaus. Some GRB light curves are generally observed with some different frequencies of radio band, and therefore the selected radio afterglows should have sufficient observational data and show plateau characteristics. Finally we obtain 27 GRBs that have a plateau phase with $0 < |\alpha_{\rm 1}| < 0.5$. The light curves of the 11 selected GRBs are exhibited in Figure 1 (other GRBs, see Figure 1 of \cite{2022ApJ...925...15L}), and the corresponding fitting parameters are listed in Table 1.

According to the fitting results, we can calculate the luminosity $\Lbz$ at the break time of the sample, using this equation:
\begin{equation}
  \Lbz = 4\pi\dl^{\rm 2}F_{\rm b}/(1+z) ,
\end{equation}
where $\dl$ is the luminosity distance. In the case of a flat ${\Lambda}$CDM cosmology model, $\dl$ can be written as
\begin{equation}\label{eq:3}
  \dl(\om,z) = (1+z)\frac{c}{H_0} \int_0^z \frac{dz}{\sqrt{\om (1+z)^3 + 1 - \om}},
\end{equation}
where $\om$ and $H_0$ are the matter density and Hubble constant at present, respectively.

When $k$-correction is included, the corresponding luminosity at the break time is
\begin{equation}\label{eq:4}
 \Lbz = 4\pi\dl^{\rm 2}F_{\rm b}(1+z)^{\rm \alpha - \beta -1},
\end{equation}
here $\alpha$ and $\beta$ are the time and frequency indices in $F \propto t^{\rm \alpha} \nu^{\rm \beta}$. We have set $\alpha = 0$ and $\beta = 1/3$ \citep{2012ApJ...746..156C}.

The prompt emission spectrum of GRBs can be described as a broken power-law known as the band function \citep{1993ApJ...413..281B}
\begin{equation}
\Phi(E) = \left \{
\begin{array}{ll}
A E^{\rm \alpha_{\gamma}} {\rm e}^{-(2 + \alpha_{\rm \gamma}) E/E_{\rm p,obs}} & E \le
\frac{\rm \alpha_{\gamma}-\beta_{\gamma}}{2 + \alpha_{\rm \gamma}}E_{\rm p,obs} \\ ~ & ~ \\
B E^{\rm \beta_{\gamma}} & {\rm otherwise,}
\end{array}
\right.\  \label{band}
\end{equation}
where $E_{\rm p,obs}$ is the peak energy in the observer's frame. $\alpha_{\rm \gamma}$ and $\beta_{\rm \gamma}$ represent the low energy photon indices and the high energy photon indices. The corresponding spectral index values of some samples are not provided in the published literature. For these samples, we have taken the typical spectral index values of GRBs, i.e., $\alpha_{\rm \gamma}=-1.0$ and $\beta_{\rm \gamma}=-2.2$.

With the energy spectrum $\Phi(E)$, the bolometric fluence $S_{\rm bolo}$ in the band of $1-10^4$ keV can be calculated by \citep{2001AJ....121.2879B}
\begin{equation}
S_{\rm bolo} = S \ {\times} \ \frac{\int_{1/(1 + z)}^{10^4/(1 +
z)}{E \Phi(E) dE}} {\int_{E_{\rm min}}^{E_{\rm max}}{E \Phi(E) dE}}
\ , \label{sbolo}
\end{equation}
where $S$ is the observed fluence, $E_{\rm min/max}$ are the detector thresholds.
Therefore, $\Eiso$ can be written as
\begin{equation}\label{eq:7}
\Eiso = 4 \pi \dl^2 S_{\rm bolo}/(1+z).
\end{equation}

Note that GRB 211106A has no redshift measurements, and here we have set $z=0.5$ as shown by \cite{2022ApJ...935L..11L}.
The full sample contains 27 GRBs in the redshift range of $0.0368 \leq z\leq 5.283$, and the values of related
parameters are listed in Table 1.

\begin{table*}[h]\footnotesize %
  \caption{The fitting results for selected radio GRBs}
  \setlength{\tabcolsep}{1.0mm}{
  \begin{center}
  \begin{tabular}{ccccccccccccc}
  \hline
  \hline
    GRB   & $z$& ${T_{\rm 90}}^b$ &${logF_{\rm b}}^c$ &${logT_{\rm b,z}^b}$ & $\alpha_{\rm 1}$  & $\alpha_{\rm 2}$ & ${logL_{\rm b,z}^d}$ & ${logE_{\rm \gamma,iso}}^e$ & ${logE_{\rm p,i}}^f$ &${Ref.}^g$  \\
  \hline
   980329$^a$	&	3.9	&	58	&	-16.64$\pm$0.04	&	5.90$\pm$0.09	&	-0.08$\pm$0.06	&	0.83$\pm$0.32	&	41.60$\pm$0.04	&	54.32$\pm$0.18	&	3.04$\pm$0.07&1,2\\
   980703$^a$	&	0.966	&	90	&	-15.99$\pm$0.11	&	5.66$\pm$0.13	&	-0.33$\pm$0.40	&	0.86$\pm$0.03	&	41.30$\pm$0.11	&	52.84$\pm$0.07	&	2.70$\pm$0.06&1,2\\
   990510	&	1.619	&	75	&	-16.70$\pm$0.02	&	5.08$\pm$0.08	&	-0.63$\pm$0.21	&	0.91$\pm$0.17	&	40.98$\pm$0.02	&	53.24$\pm$0.02	&	2.55$\pm$0.03&3,3\\
   991208	&	0.706	&	60	&	-16.43$\pm$0.20	&	6.30$\pm$0.26	&	0.35$\pm$0.18	&	1.22$\pm$0.35	&	40.61$\pm$0.20	&	53.36$\pm$0.04	&	2.50$\pm$0.04&2,2\\
   000418	&	1.119	&	30	&	-16.09$\pm$0.05	&	5.94$\pm$0.07	&	-0.56$\pm$0.36	&	1.30$\pm$0.14	&	41.31$\pm$0.05	&	52.98$\pm$0.08	&	2.45$\pm$0.03&2,2\\
   000926$^a$	&	2.039	&	25	&	-16.78$\pm$0.05	&	5.53$\pm$0.05	&	-0.22$\pm$0.20	&	0.45$\pm$0.15	&	41.06$\pm$0.05	&	53.43$\pm$0.11	&	2.49$\pm$0.03&1,2\\
   010222$^a$	&	1.477	&	170	&	-17.38$\pm$0.33	&	6.41$\pm$0.28	&	0.06$\pm$0.19	&	1.33$\pm$0.64	&	40.23$\pm$0.33	&	54.12$\pm$0.13	&	2.88$\pm$0.02&1,2\\
   011030$^a$	&	3	&	1500	&	-16.89$\pm$0.05	&	5.72$\pm$0.07	&	-0.21$\pm$0.20	&	0.91$\pm$0.25	&	41.20$\pm$0.05	&	51.69$\pm$0.50	&	1.56$\pm$0.10&1,4\\
   020903$^a$	&	0.25	&	13	&	-16.79$\pm$0.29	&	7.06$\pm$0.10	&	0.45$\pm$0.38	&	7.29$\pm$11.20	&	39.36$\pm$0.29	&	49.36$\pm$0.12	&	0.53$\pm$0.23&1,2\\
   021004$^a$	&	2.33	&	50	&	-16.27$\pm$0.04	&	5.92$\pm$0.04	&	-0.12$\pm$0.08	&	1.30$\pm$0.14	&	41.66$\pm$0.04	&	52.58$\pm$0.14	&	2.42$\pm$0.19&1,2\\
   030329$^a$	&	0.168	&	63	&	-13.65$\pm$0.02	&	5.87$\pm$0.02	&	-0.09$\pm$0.05	&	1.84$\pm$0.16	&	42.15$\pm$0.02	&	52.26$\pm$0.13	&	2.00$\pm$0.19&1,2\\
   031203	&	0.105	&	30	&	-16.50$\pm$0.03	&	6.67$\pm$0.31	&	-0.30$\pm$0.17	&	0.58$\pm$0.61	&	38.89$\pm$0.03	&	50.00$\pm$0.17	&	2.20$\pm$0.14&1,3\\
   050713B$^a$	&	0.55	&	54.2	&	-16.68$\pm$0.06	&	6.12$\pm$0.07	&	0.25$\pm$0.14	&	1.90$\pm$1.05	&	40.15$\pm$0.06	&	49.71$\pm$0.41	&	2.25$\pm$0.13&1,5\\
   070125	&	1.548	&	60	&	-16.35$\pm$0.03	&	6.12$\pm$0.11	&	-0.26$\pm$0.10	&	0.95$\pm$0.19	&	41.30$\pm$0.03	&	53.92$\pm$0.04	&	2.97$\pm$0.07&2,2\\
   070612A$^a$	&	0.617	&	368.8	&	-16.07$\pm$0.03	&	6.75$\pm$0.05	&	-0.21$\pm$0.07	&	1.52$\pm$0.45	&	40.85$\pm$0.03	&	51.96$\pm$0.13	&	2.35$\pm$0.00&1,6\\
   071003$^a$	&	1.1	&	150	&	-16.40$\pm$0.12	&	5.50$\pm$0.20	&	-0.12$\pm$0.31	&	0.68$\pm$0.25	&	40.99$\pm$0.12	&	51.55$\pm$0.03	&	3.32$\pm$0.06&1,2\\
   090323	&	3.57	&	133	&	-16.81$\pm$0.09	&	5.66$\pm$0.14	&	-0.20$\pm$0.27	&	2.70$\pm$2.54	&	41.38$\pm$0.09	&	54.64$\pm$0.05	&	3.28$\pm$0.08&2,2\\
   111215A$^a$	&	2.06	&	796	&	-15.40$\pm$0.02	&	5.57$\pm$0.02	&	0.15$\pm$0.03	&	1.08$\pm$0.08	&	42.44$\pm$0.02	&	53.29$\pm$0.12	&	2.43$\pm$0.08&1,6\\
   120326A$^a$	&	1.798	&	69.6	&	-16.40$\pm$0.17	&	6.14$\pm$0.27	&	-0.12$\pm$0.30	&	2.04$\pm$2.92	&	41.35$\pm$0.17	&	52.51$\pm$0.04	&	2.18$\pm$0.04&1,2\\
   140304A$^a$	&	5.283	&	31.2	&	-16.56$\pm$0.04	&	4.93$\pm$0.07	&	-0.11$\pm$0.11	&	0.89$\pm$0.16	&	41.85$\pm$0.04	&	53.14$\pm$0.03	&	2.94$\pm$0.05&1,3\\
   141121A$^a$	&	1.47	&	33.2	&	-16.72$\pm$0.08	&	5.90$\pm$0.11	&	0.28$\pm$0.18	&	1.10$\pm$0.71	&	40.89$\pm$0.08	&	50.83$\pm$0.16	&	2.32$\pm$0.14&1,7\\
   160623A$^h$	&	0.367	&	...	&	-13.59$\pm$0.46	&	5.15$\pm$0.67	&	0.23$\pm$0.91	&	1.10$\pm$0.54	&	42.89$\pm$0.46	&	53.40$\pm$0.01	&	2.83$\pm$0.01&8,8\\
   171010A$^a$	&	0.3285	&	104	&	-15.40$\pm$0.02	&	5.40$\pm$0.03	&	-0.11$\pm$0.09	&	1.11$\pm$0.05	&	40.99$\pm$0.02	&	53.26$\pm$0.01	&	2.36$\pm$0.02&1,3\\
   171205A	&	0.0368	&	189.4	&	-13.01$\pm$0.06	&	5.41$\pm$0.08	&	0.01$\pm$0.21	&	1.46$\pm$0.18	&	41.47$\pm$0.06	&	49.34$\pm$0.13	&	2.10$\pm$0.13&3,3\\
   190114C	&	0.425	&	120	&	-15.75$\pm$0.02	&	5.62$\pm$0.03	&	-0.19$\pm$0.06	&	1.11$\pm$0.04	&	40.86$\pm$0.02	&	53.43$\pm$0.00	&	2.97$\pm$0.00&3,3\\
   191221B & 1.148  &  48 &  -14.26$\pm$0.01  &  4.93$\pm$0.02  &  -0.26$\pm$0.02  &  1.15$\pm$0.11  &  43.16$\pm$0.01 & 53.56$\pm$0.05 &  2.91$\pm$0.03& 9,9\\
   211106A	&	0.5	&	1.71	&	-17.13$\pm$0.03	&	6.52$\pm$0.04	&	-0.28$\pm$0.10	&	2.51$\pm$0.76	&	39.62$\pm$0.03	&	51.04$\pm$0.04	&	2.66$\pm$0.09&10,10\\
 \hline
\end{tabular}
\end{center}}
\tablecomments{\\$^{a}$ GRBs from \cite{2022ApJ...925...15L}\\
$^{b}$  In units of seconds.\\
$^c$ In units of $\rm erg$ $\rm cm^{\rm -2}$ $\rm s^{\rm -1}$. \\
$^d$ In units of $\rm erg$ $\rm s^{\rm -1}$. \\
$^e$ In units of $\rm erg$. \\
$^f$ In units of $\rm keV$. \\
$^g$ References for $E_{\rm \gamma,iso}$ and${E_{\rm p,i}}$. \\
$^h$ The $T_{90}$ of GRB 160623A is not available.}
\tablerefs{(1) \cite{2022ApJ...925...15L}; (2)\cite{2017A&A...598A.112D}; (3)\cite{2020MNRAS.492.1919M}; (4)\cite{2006A&A...455..413G}; (5)\cite{2019ApJ...883...97Z}; (6)\cite{2018ApJ...866...97B}; (7)\cite{2020ApJ...900..112Z}; (8)\cite{2020ApJ...891L..15C};(9)\cite{2023NatAs...7...80U}; (10)\cite{2022ApJ...935L..11L}.}
\end{table*}

\begin{figure*}[htbp!]
\center
\includegraphics[angle=0,width=0.30\textwidth]{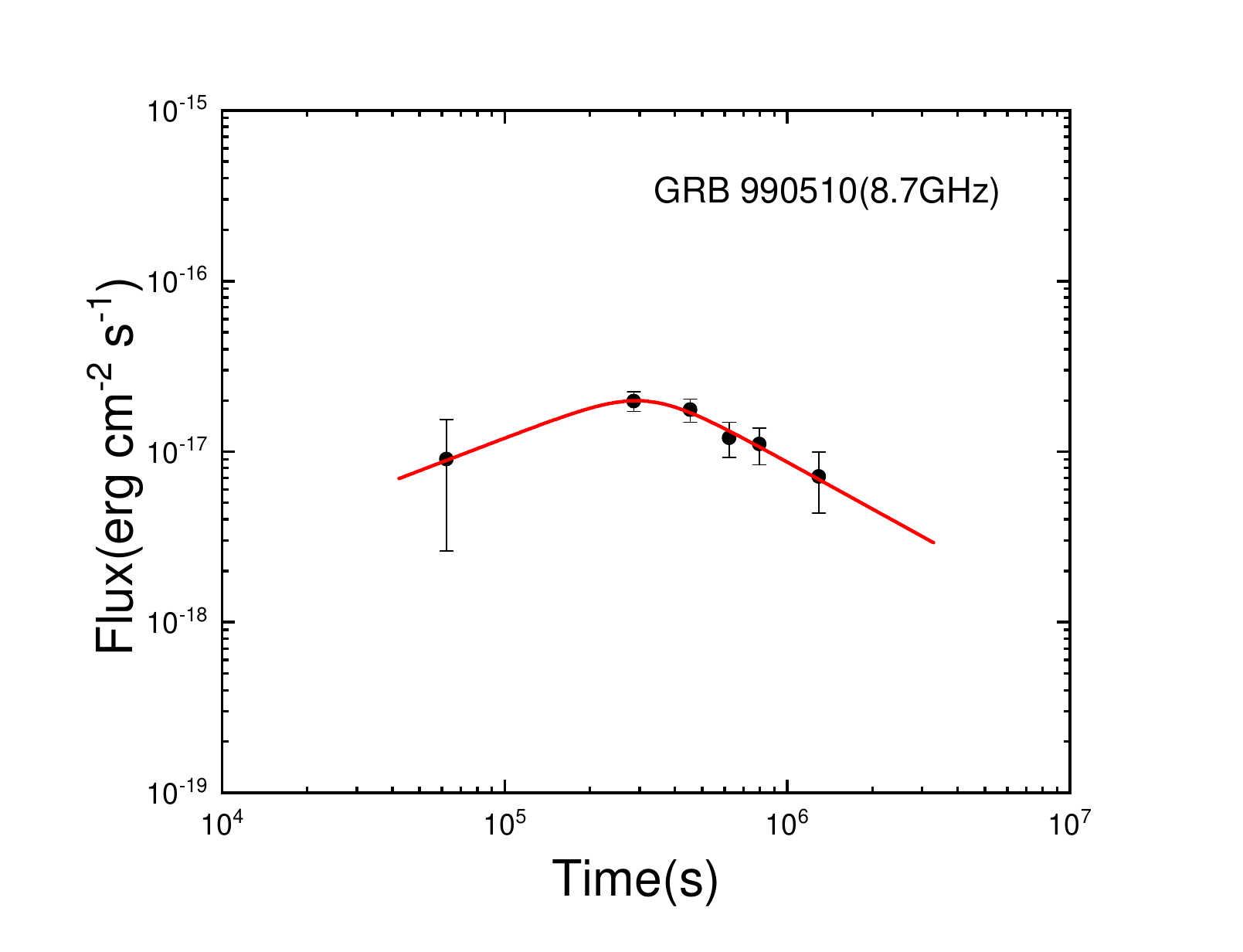}\includegraphics[angle=0,width=0.30\textwidth]{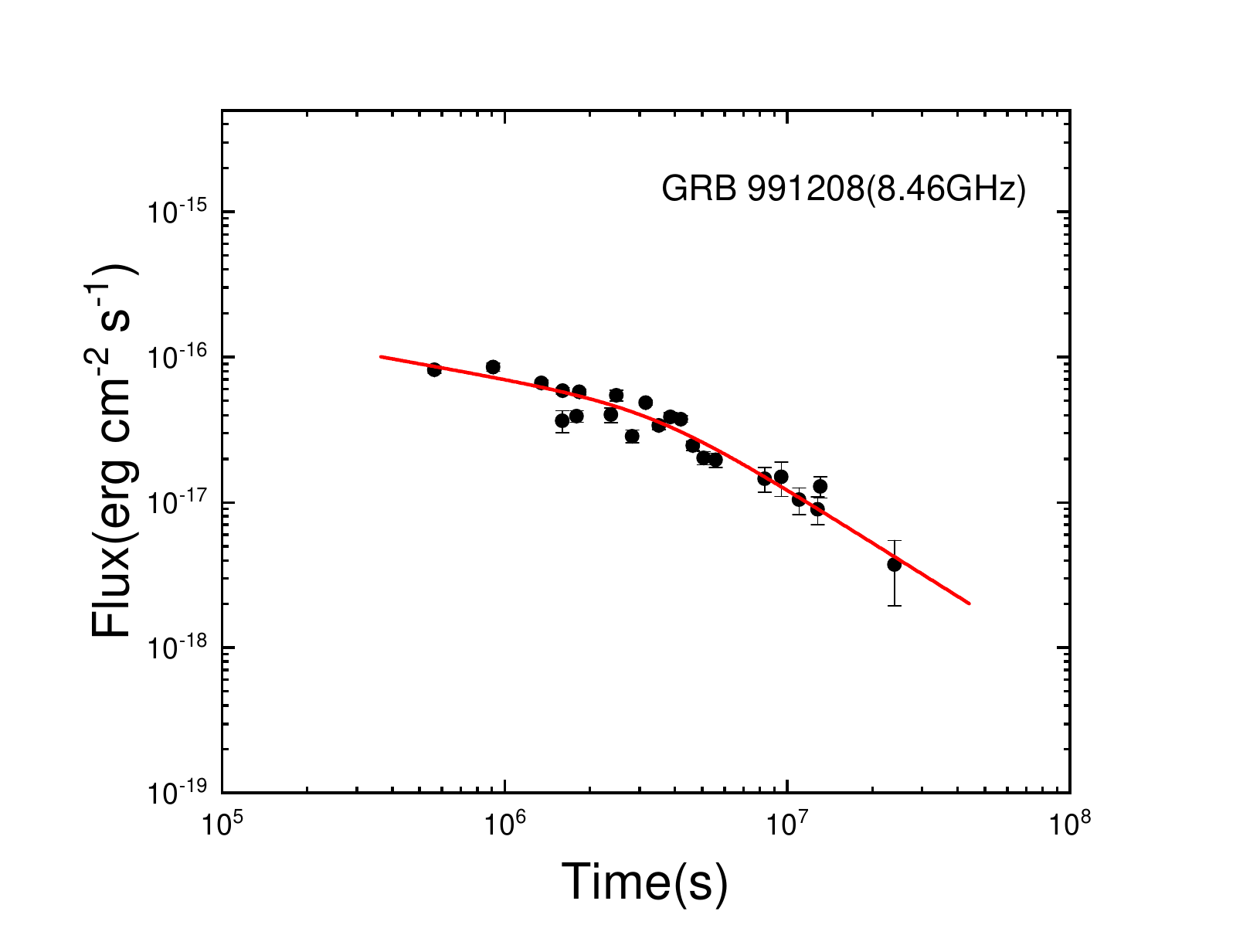}\includegraphics[angle=0,width=0.30\textwidth]{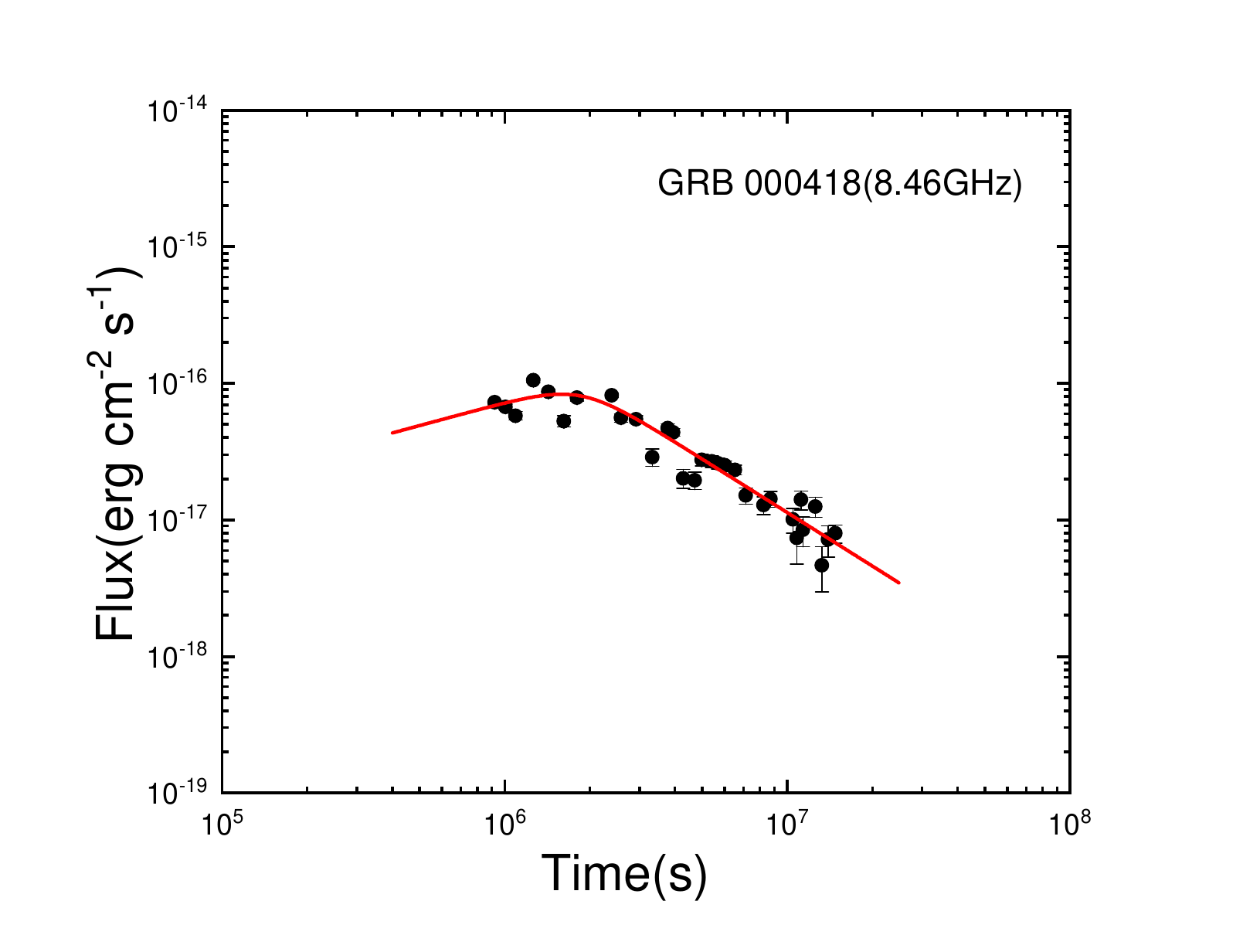}
\includegraphics[angle=0,width=0.30\textwidth]{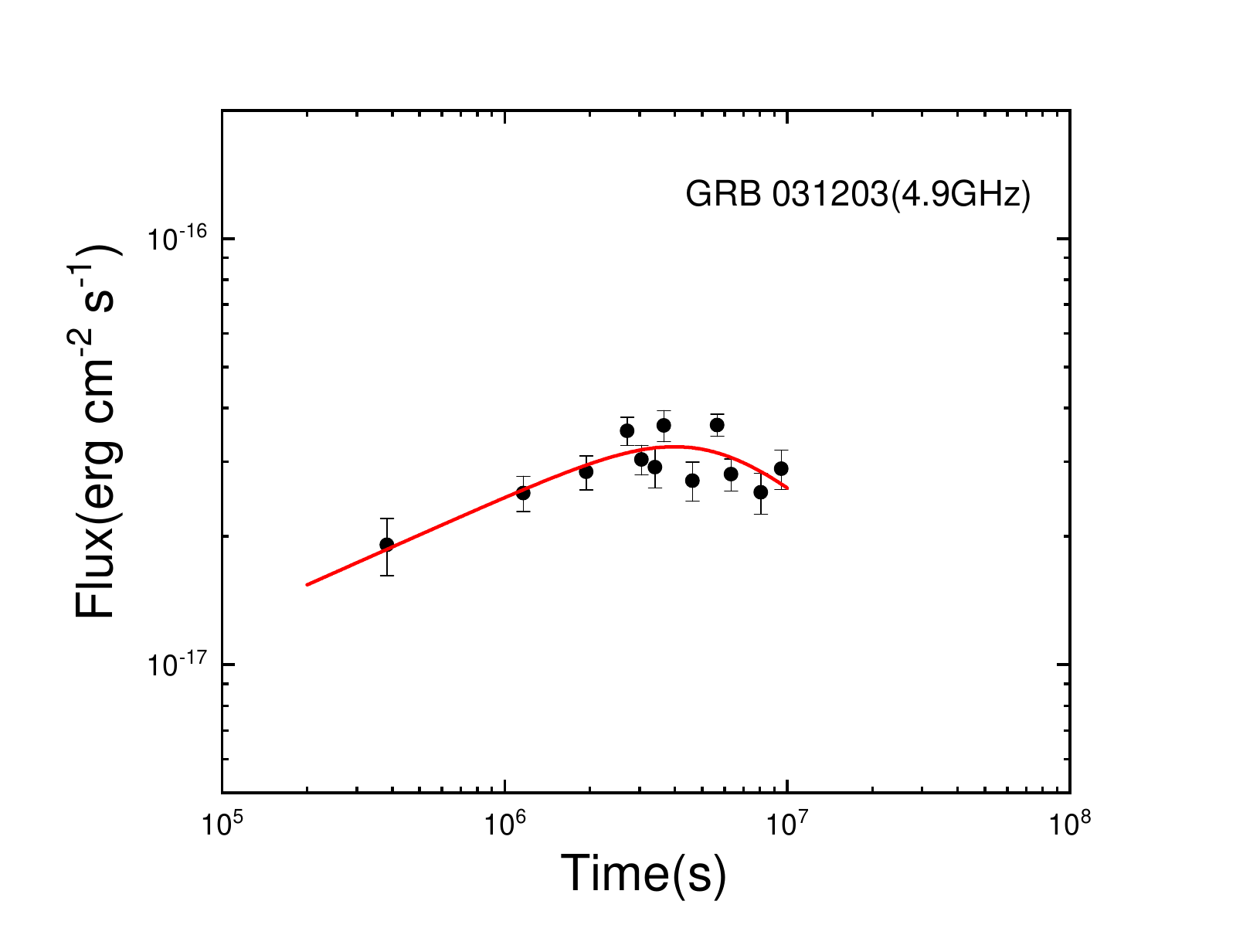}\includegraphics[angle=0,width=0.30\textwidth]{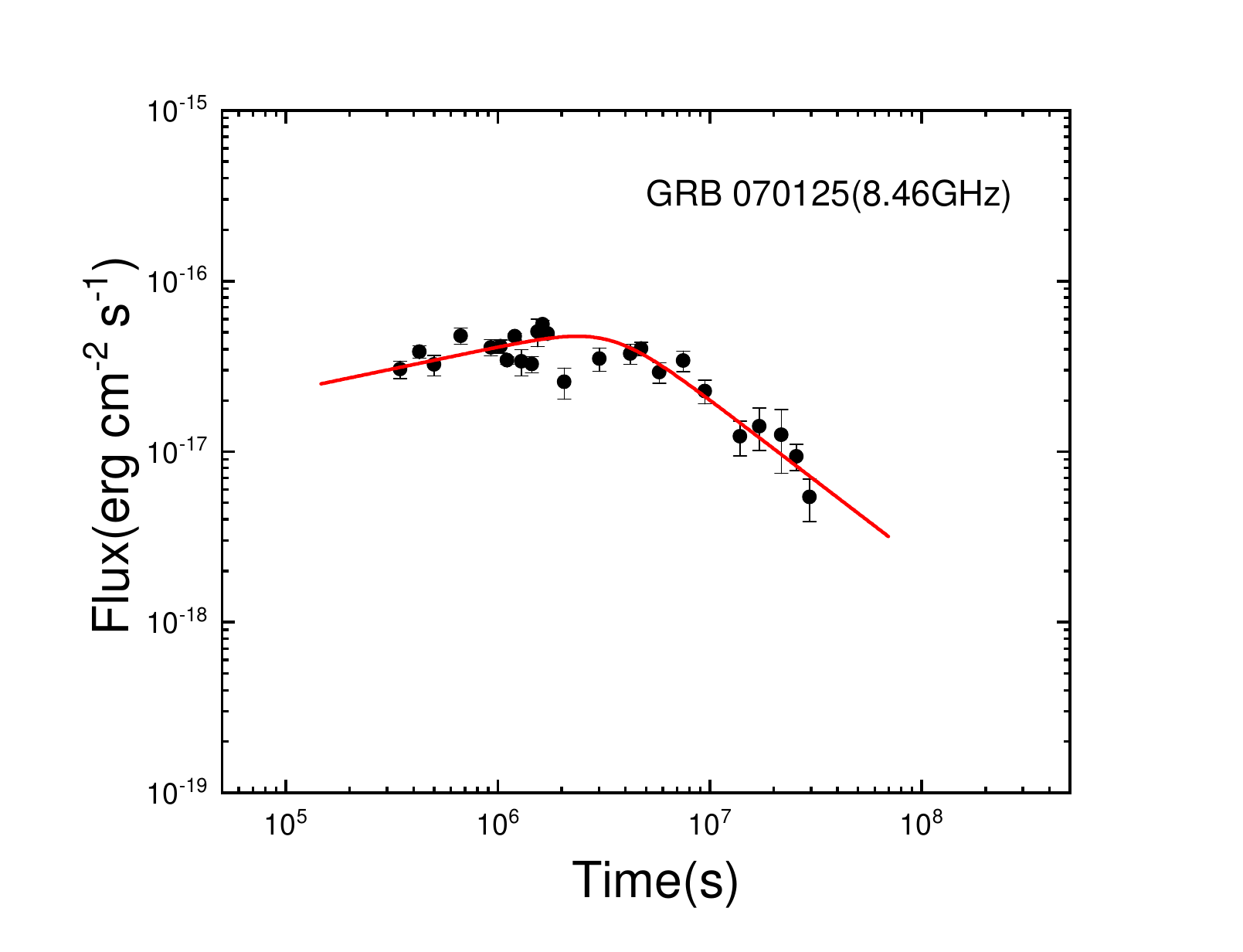}\includegraphics[angle=0,width=0.30\textwidth]{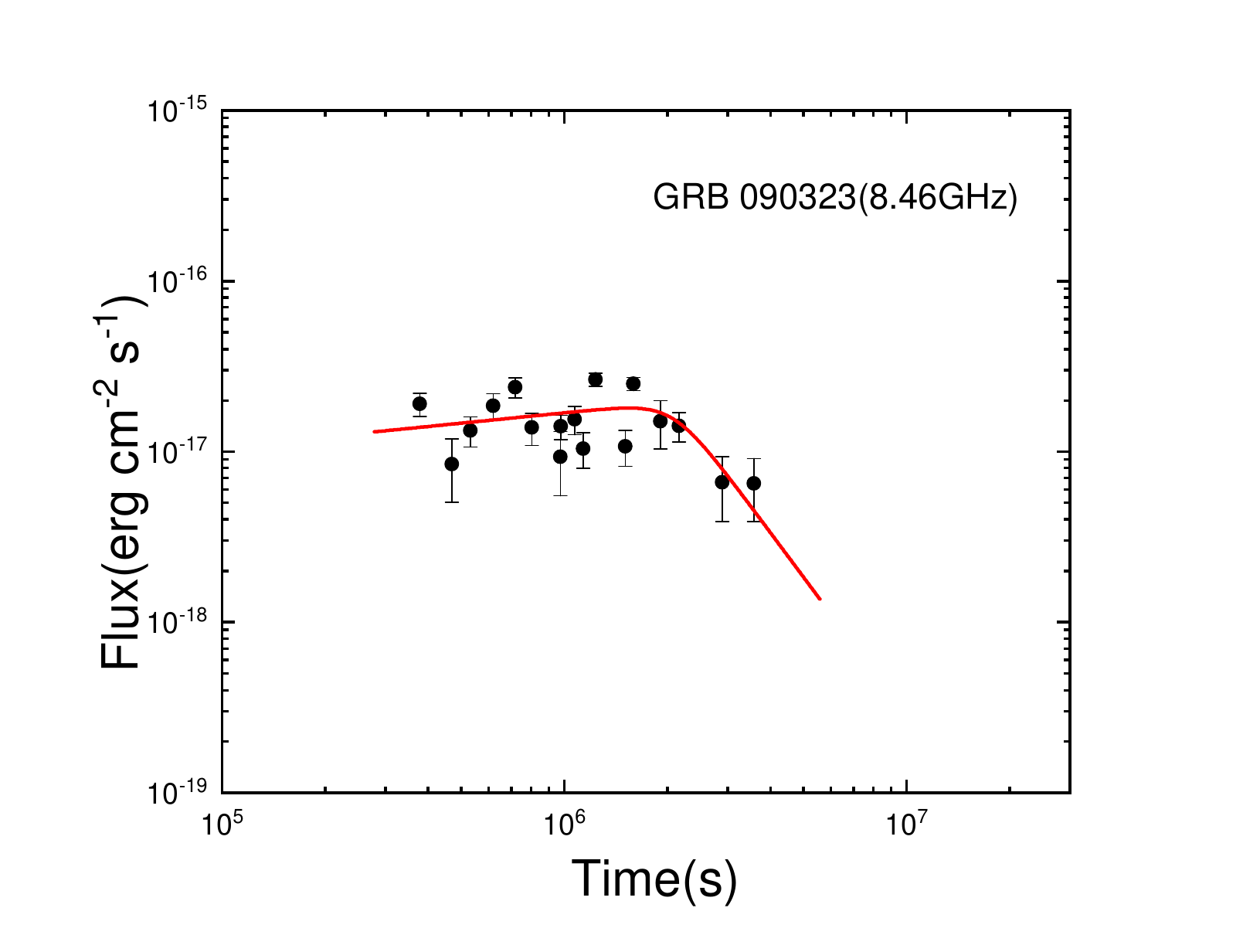}
\includegraphics[angle=0,width=0.30\textwidth]{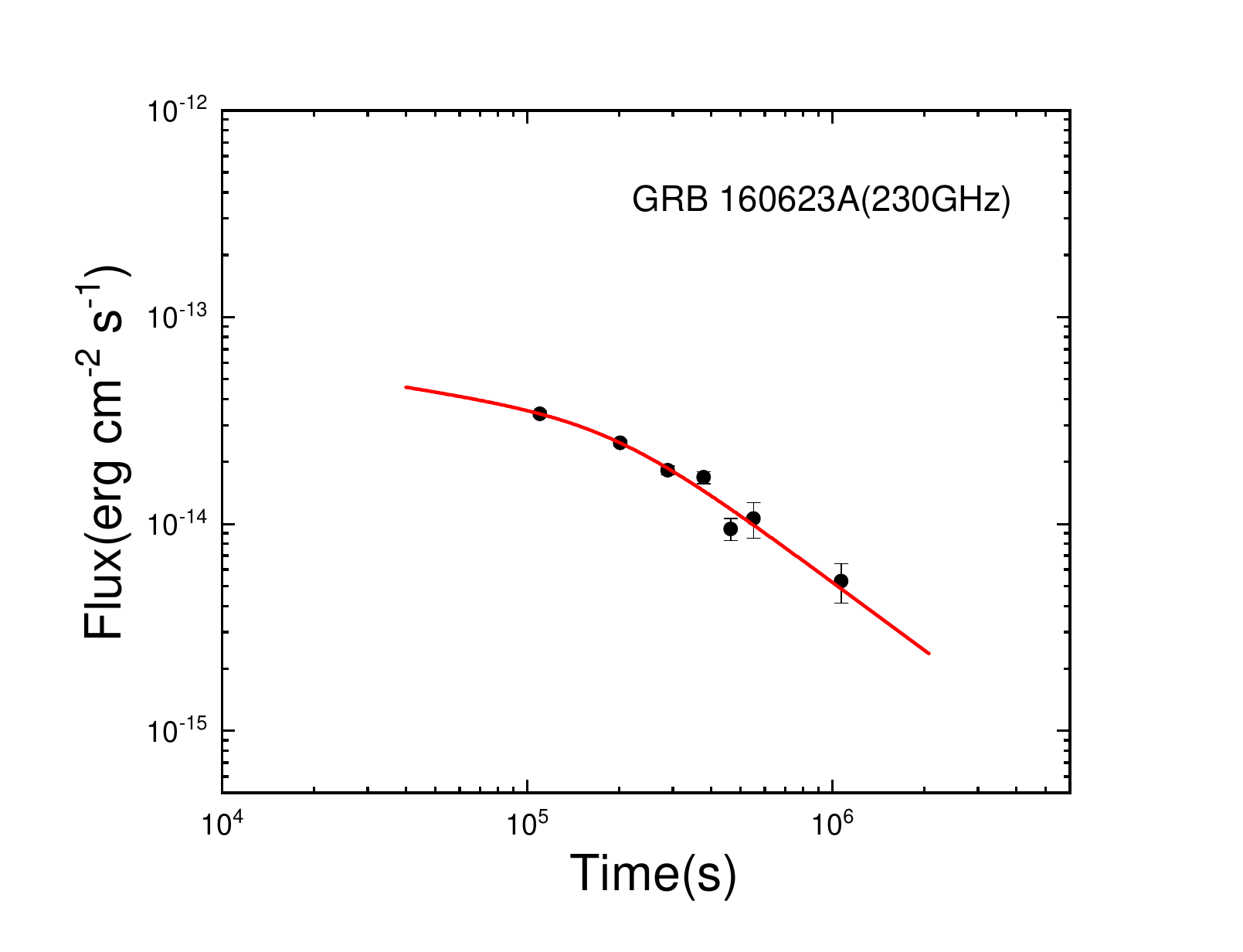}\includegraphics[angle=0,width=0.30\textwidth]{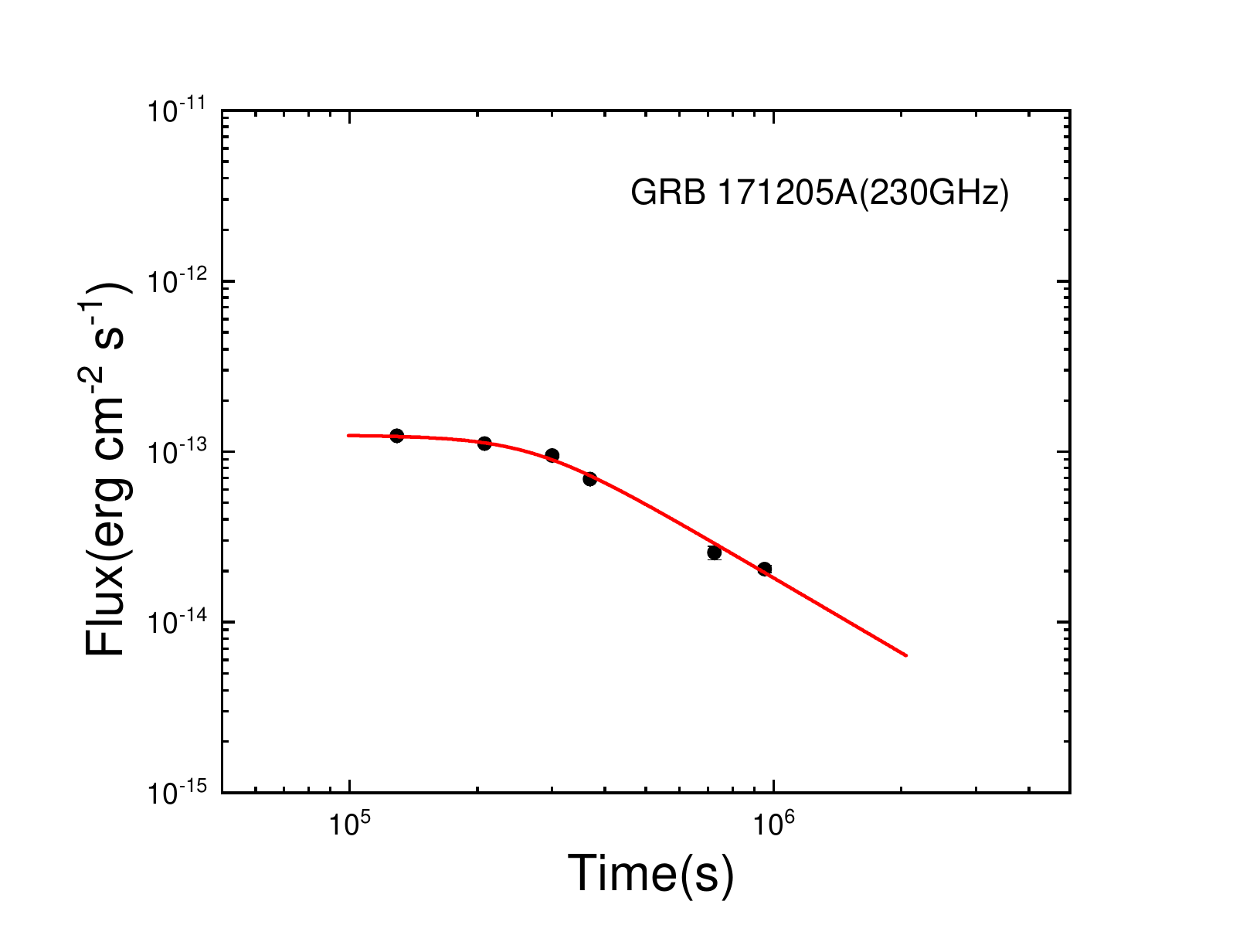}\includegraphics[angle=0,width=0.30\textwidth]{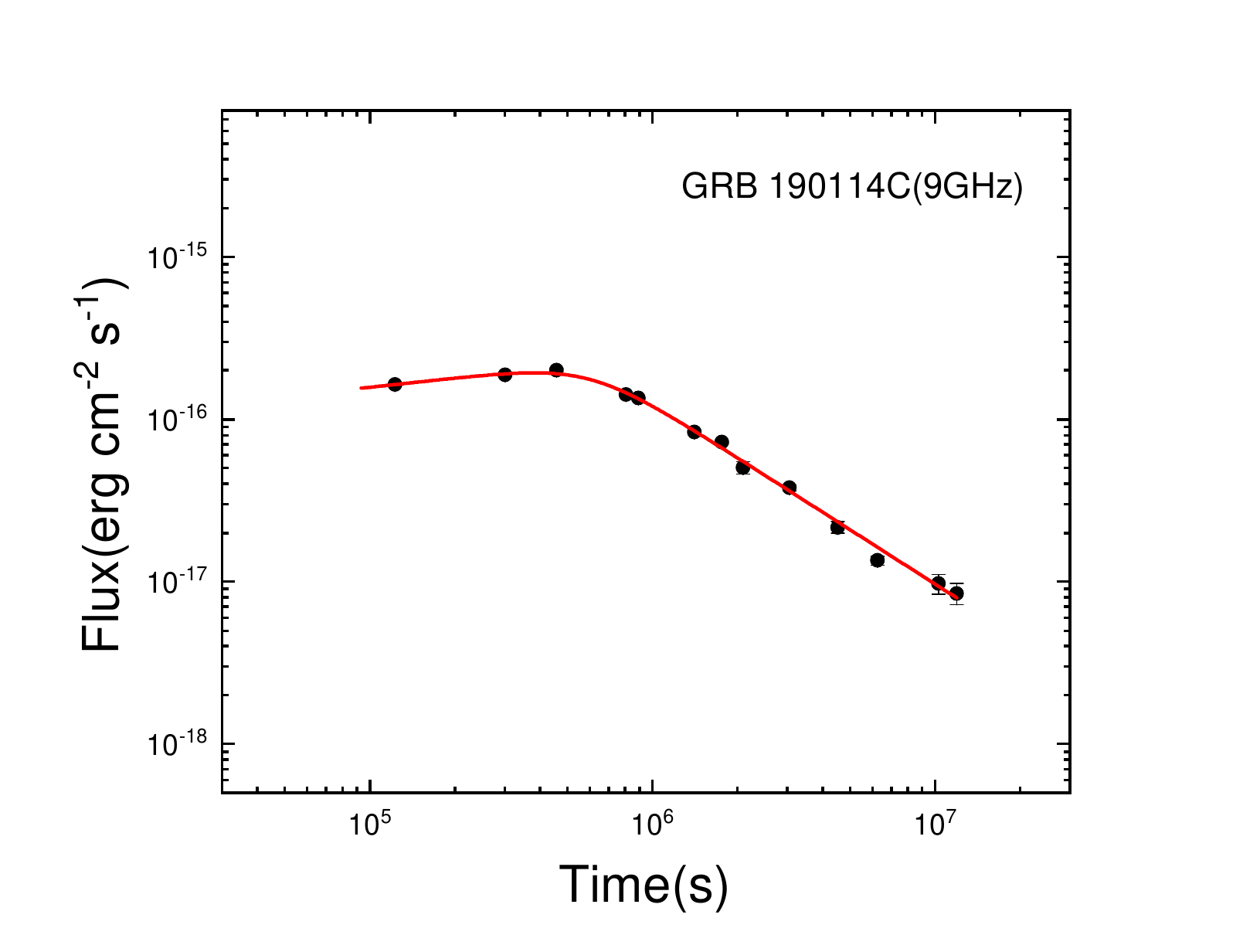}
\includegraphics[angle=0,width=0.30\textwidth]{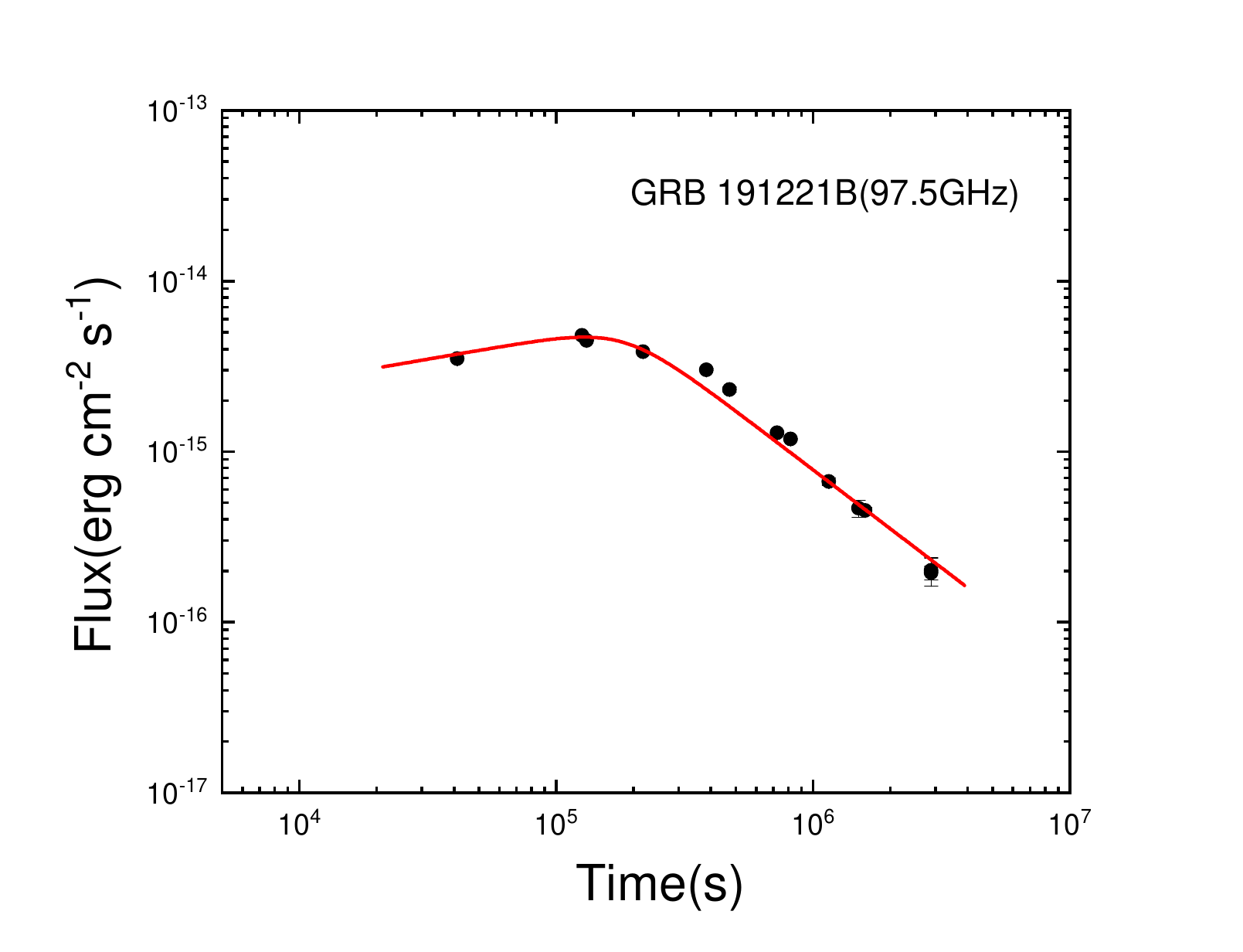}\includegraphics[angle=0,width=0.30\textwidth]{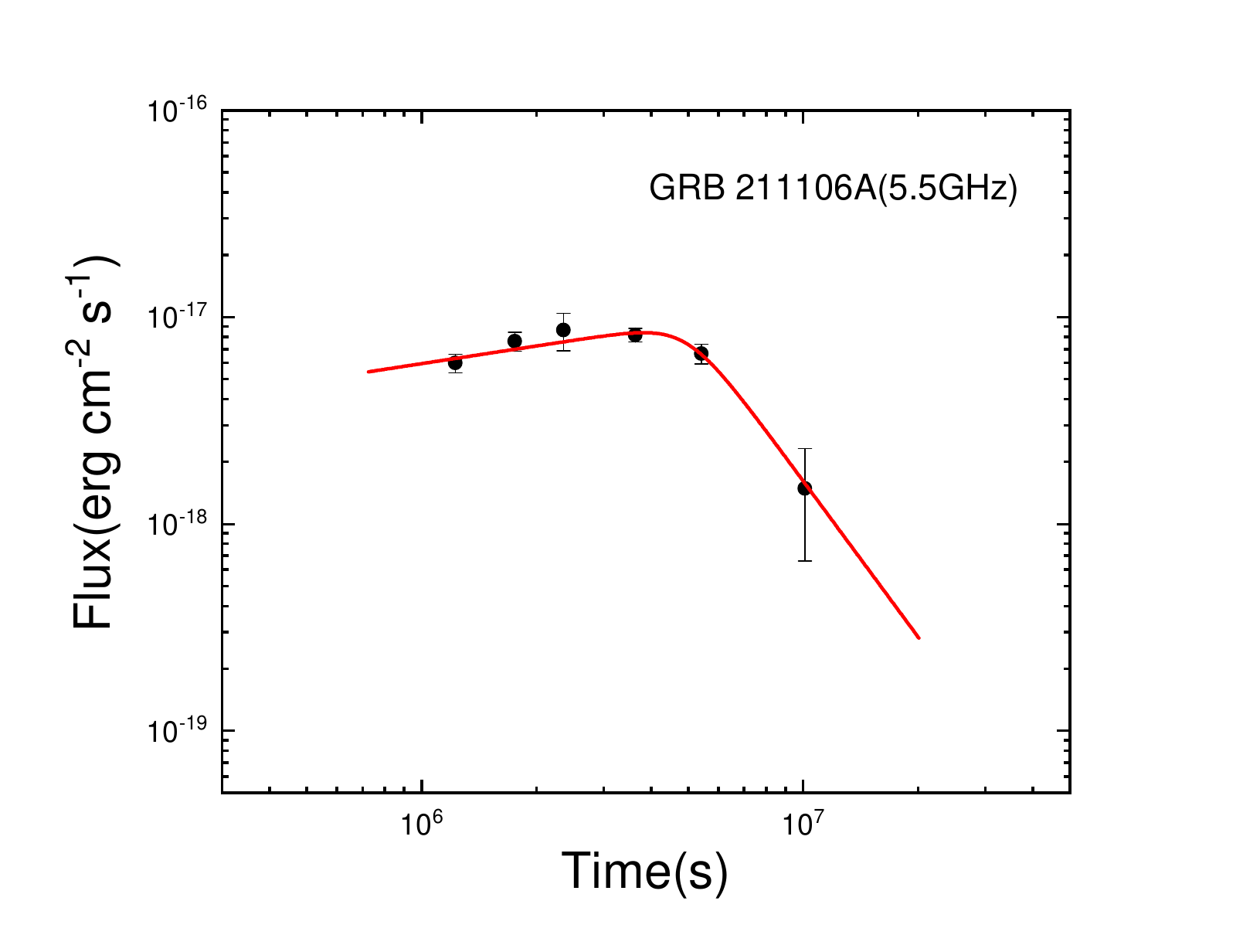}
\caption{ Fitting results of radio light curves with plateau components. We have used a smooth broken power-law function to fit these curves. The black dots represent the radio data, and the solid red lines represent the best fit to the radio data.}
\end{figure*}

\section{Fitting Results}

\subsection{Two-parameter Correlations}
The correlations between various parameter pairs of GRBs in the rest frame have been widely studied, such as Dainotti relation. We parameterize the $\Lbz$-$\Tbz$ correlation as follows:
\begin{equation}
	\log{\frac{\Lbz}{10^{40}~\rm{erg~s^{-1}}}}=b + k~\log{\frac{\Tbz}{10^{5}~\rm{s}}}.
\end{equation}

The Dainotti relation can be expressed as $y = kx + b$, and the Markov Chain Monte Calo (MCMC) technique is applied to obtain the best fit values of related parameters. The likelihood function in this section can be written as \citep{2005physics..11182D}
\begin{eqnarray}
	\mathcal{L}(k,b,\ext ) &\propto&
	\prod\limits_i {\frac{1}{{\sqrt {\sigma ^2 _{{\mathop{\rm int}} }  +
					\sigma ^2 _{y_{ i} }  + k^2 \sigma ^2 _{x_{ i} } } }}} \nonumber \\
	&\times&
	\exp \left[ - \frac{{(y_{ i} - kx_{ i} - b )^2 }}{{2(\sigma ^2 _{{\mathop{\rm
						int}} }  + \sigma ^2 _{y_{ i} }  + k^2 \sigma ^2 _{x_{i} } )}}\right],
\end{eqnarray}
where $\sigma_{\rm int}$ is the intrinsic scatter. Figure 2 shows Dainotti relation for selected sample, where the observed GRB data points are distributed tightly along the red line called the best-fit line. The best fitting results are $k = -1.20\pm 0.24$ and $b = 2.13 \pm 0.24$ with intrinsic scatter $\sigma_{\rm int} = 0.66\pm0.09$. Spearman coefficient and a test $p$-value for correlation hypothesis are adopted to discover a more obvious correlation.
When the $p$-value is less than 0.05, it indicates that the possibility of the correlation being true is very high. Meanwhile, the closer the absolute value of Spearman coefficient is to 1, the tighter of the correlation is. Here we calculated that the Spearman correlation coefficient $\rho$ is -0.58, with a chance probability of $p = 2\times10^{-3}$, indicating that the correlation is significant. The Pearson coefficient is $r = -0.70$.

It could be distinctly indicated that $\Lbz$ and $\Tbz$ present a tight negative correlation, $\Lbz \propto \Tbz^{-1.20\pm 0.24}$. This result for radio plateaus is similar to that of X-ray afterglows with antic-correlation $\Lbz \propto \Tbz^{-1.06\pm0.27}$ reported by \cite{2010ApJ...722L.215D}. We found that the longer time of plateau associates with the dimmer break luminosity. The break time of radio plateau samples we selected is later than that of X-ray and optical plateaus, and the break luminosity is several orders of magnitude lower than that of X-ray and optical afterglow plateaus.

The Efron-Petrosian method can be used to correct variables and study the intrinsic nature of correlations, which is able to overcome the challenge of selection effects and redshift evolution \citep{1992ApJ...399..345E}. \cite{2021Galax...9...95D} used this method to correct some important variables of GRB radio afterglows, including $\Eiso$ and $T_{90}$ in the rest frame for 80 GRBs, and $\Lbz$ and $\Tbz$ for a subsample of 18 GRBs with plateau phase. In addition, using the selected 14 radio afterglows with plateau phase, \cite{2022ApJ...925...15L} summarized the Dainotti relation as $\Lbz \propto \Tbz^{-2.34\pm0.66}$. Under this circumstance, the Dainotti correlation result of the radio plateau corrected by the Efron-Petrosian method $\Lbz \propto \Tbz^{-0.26\pm0.71}$ was obtained.

\begin{figure*}[htbp!]
\center
\includegraphics[angle=0,width=0.45\textwidth]{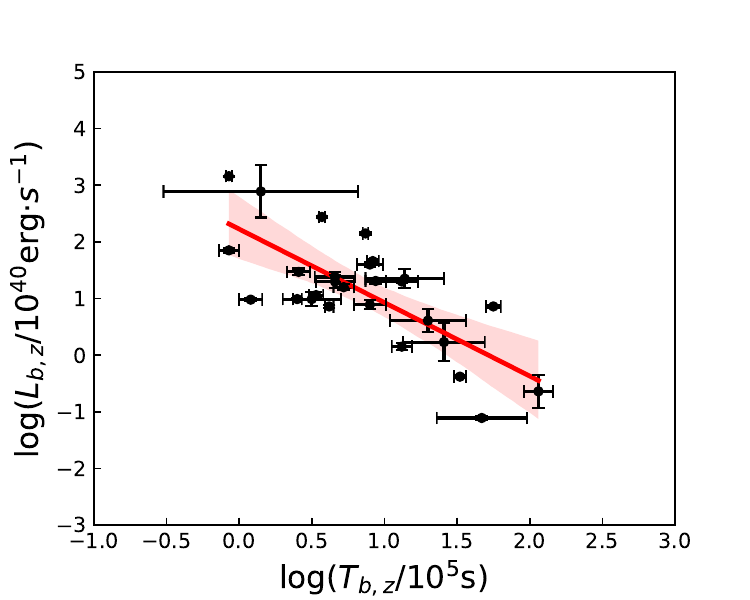}
\caption{ Logarithmic plots of two-parameter correlations: $\Lbz$-$\Tbz$. The red shadow region is 95$\%$ confidence levels.}
\end{figure*}

\subsection{Three-parameter Correlations}
A tight three-parameter correlation, called the $\LTE$, has been mentioned by \cite{2012A&A...538A.134X}, by studying a group of X-ray afterglows with plateau phase \citep{2023ApJ...943..126D}. In this paper, we try to use radio plateaus to search for possible multivariate correlations, depending on the correlations of $\Lbz$-$\Tbz$. We gather the isotropic energy of each GRB with a radio plateau, and then plan to investigate whether the $\LTE$ correlation still exists for our radio afterglow sample. The correlation can be written as:
\begin{equation}\label{eq:10}
	\log{\frac{\Lbz}{10^{40}\rm{erg~s^{-1}}}}=a + b~\log{\frac{\Tbz}{10^{5}~\rm{s}}}
      + c~\log{\frac{\Eiso}{10^{52}~\rm{erg}}},
\end{equation}
where $a$, $b$ and $c$ are the coefficients to be measured, which can be obtained by fitting the observational data. In this equation, $\Lbz$ is approximated as a function of break time $\Tbz$ and isotropic equivalent energy $\Eiso$. Here, $a$ is a constant.

Similarly, MCMC algorithm is used to obtain the best fit values of related parameters. The likelihood function for three-parameter correlations can be written as \citep{2005physics..11182D}
\begin{eqnarray}\label{eq:11}
	\mathcal{L}(a,b,c,\ext)\propto
    \prod\limits_i {\frac{1}{\sqrt{ \ext^{2} + \sigma_{y_{i}}^{2} + b^{2}\sigma_{x_{1,i}}^{2} + c^{2}\sigma_{x_{2,i}}^{2} } }}\nonumber\\
    \times\exp \left[ -\frac{( y_{i} - a - bx_{1,i} - cx_{2,i})^{2} }{ 2( \ext^{2} + \sigma_{y_{i}}^{2}
     + b^{2}\sigma_{x_{1,i}}^{2} + c^{2}\sigma_{x_{2,i}}^{2}) }\right],
\end{eqnarray}
where $x_1 = \log(\Tbz/{10^5s})$, $x_2 = \log(\Eiso/{10^{52}~erg})$ and $y = \log(\Lbz/{10^{40}erg~s^{-1}})$. $\sigma_{x_{1,i}}$, $\sigma_{x_{2,i}}$ and $\sigma_{y_i}$ are errors of $x_1$, $x_2$ and $y$, respectively. We obtain the best fit values of $a = 1.90\pm0.26$, $b = -1.01\pm0.24$, $c = 0.18\pm0.09$ and $\sigma_{\rm int} = 0.63\pm0.09$. Therefore, equation (\ref{eq:10}) can be rewritten as
\begin{eqnarray}
	\log{\frac{\Lbz}{10^{40}\rm{erg~s^{-1}}}}&=&(1.90\pm0.26) + (-1.01\pm0.24)\log{\frac{\Tbz}{10^{5}\rm{s}}}
    \nonumber \\
    &+& (0.18\pm0.09)\log{\frac{\Eiso}{10^{52}\rm{erg}}}.
\end{eqnarray}

We also tested the hypothesis of this three-parameter linear regression models. Similarly, there is a high probability that the model is true when $p$-value is smaller than 0.05. According to our calculations, the Spearman correlation coefficient is $\rho$ = 0.61, with $p = 1\times10^{-3}$, and the Pearson coefficient is $r = 0.75$. The best fitting results are illustrated in Figure 3.

The Amati relation \citep{2002A&A...390...81A} describes the tight correlation between $\Eiso$ and $\Epi$. Combined with the previous work, this means that it should be feasible to explore the correlation between the parameters $\Lbz$, $\Tbz$ and $\Epi$. We have selected the rest frame peak energy $\Epi = E_{\rm p,obs}(1 + z)$ from the published literature. These data are also summarized in Table 1. Using the likelihood function (\ref{eq:11}) instead of $x_1 = \log(\Tbz/{\rm 10^5s})$, $x_2 = \log(\Epi/{\rm 10^2keV})$ and $y = \log(\Lbz/{\rm 10^{40}erg~s^{-1}})$, we obtained the correlation as follows
\begin{eqnarray}
	\log{\frac{\Lbz}{10^{40}\rm{erg~s^{-1}}}}&=&(2.10\pm0.34)+(-1.18\pm0.27)\log{\frac{\Tbz}{10^{5}\rm{s}}}
    \nonumber \\
      &+&(0.05\pm0.28)\log{\frac{\Epi}{10^{2}\rm keV}},
\end{eqnarray}
where $\sigma_{\rm int} = 0.67\pm0.09$. The Pearson coefficient is $r = 0.70$, and the Spearman correlation coefficient is $\rho = 0.59$ with $p=1.6\times10^{-3}$. These correlation coefficient indicate that there are many possibilities for $\LTEpi$ correlation to be established. The best fitting result is shown in Figure 3.

It is of interest to note that the $\LTE$ and $\LTEpi$ correlations of the radio plateau sample still exists. The results are shown in Table 2, indicating $\Lbz \propto \Tbz^{-1.01 \pm 0.24} \Eiso^{0.18 \pm 0.09}$, $\Lbz \propto \Tbz^{-1.18\pm0.27} \Epi^{0.05\pm0.28}$. Note that the relations are not completely consistent with the results of X-ray and optical afterglows, implying that the radio plateaus may have different origins. The results of these two correlations appear a relatively tight relationship between luminosity and energy, which can be used as a standard candle to constrain cosmological parameters. We will cover these issues in the next section.
\begin{figure*}[htbp!]
\center
\includegraphics[angle=0,width=0.45\textwidth]{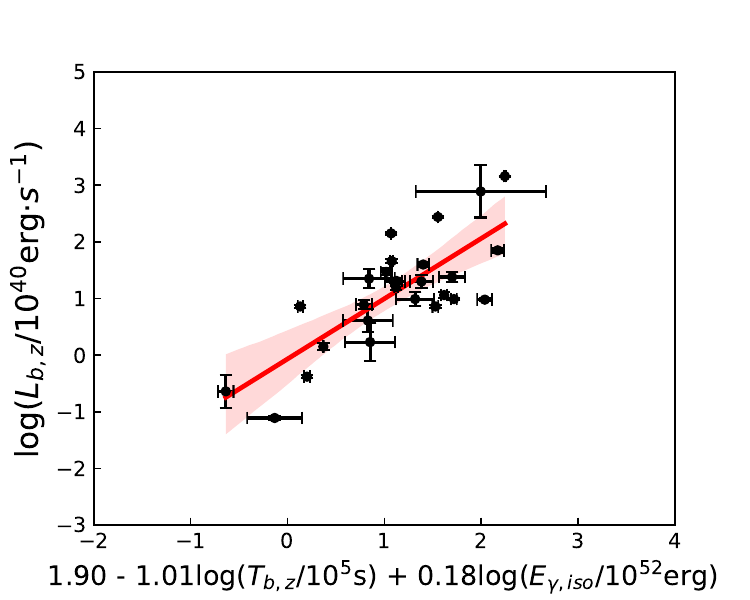}\includegraphics[angle=0,width=0.45\textwidth]{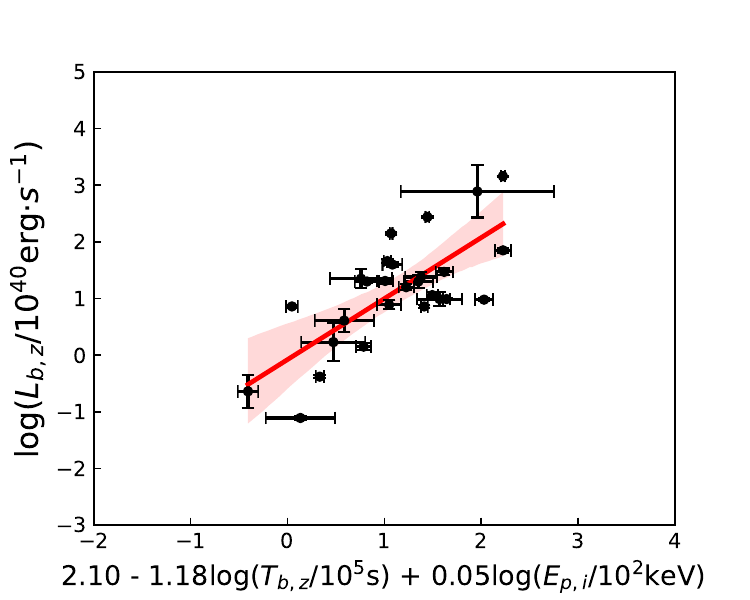}
\caption{ The best fit for two three-parameter correlations: $\Lbz$-$\Tbz$-$\Eiso$ and $\Lbz$-$\Tbz$-$\Epi$. Here, $\Lbz$ is the luminosity at the break time of the plateau, $\Tbz$ is the corresponding break time in the rest frame. $\Eiso$ and $\Epi$ are the isotropic and peak gamma-ray energy of the prompt emission, respectively. The red shadow region is 95$\%$ confidence levels.}
\end{figure*}

\section{CONSTRAINING COSMOLOGICAL MODELS}

\subsection{Calibrating GRBs correlations }

In the previous section, we have deduced various correlations by assuming constant values for cosmological parameters.
It has been recommend that long GRBs have a advantage of expanding Hubble diagram to high redshifts \citep{2001ApJ...562L..55F,2004ApJ...612L.101D,2005ApJ...633..611L,2007ApJ...660...16S}. In order to use our selected GRB sample to do this issue, one ought to calibrate the correlations of GRBs. Moreover, the circularity problem should be confronted when empirical luminosity correlation of GRBs is used to constrain cosmological parameters. Several methods have been proposed to overcome this problem (\citealt{2008A&A...490...31C,2008MNRAS.391L...1K,2011MNRAS.415.3423W,2016A&A...585A..68W,2017NewAR..77...23D,2018AdAst2018E...1D,2018PASP..130e1001D,2019gbcc.book.....D,2021MNRAS.507..730H,2022MNRAS.516.2575J,2022ApJ...941...84L};\citealt{2022MNRAS.510.2928C}b).

Different authors have taken different calibration methods to avoid this problem in order to obtain the intrinsic GRB correlations. For example, \cite{2008MNRAS.391..577A} and \cite{2013MNRAS.436...82D}a used a simultaneous fitting method. The Efron-Petrosian method was applied to obtain the intrinsic properties of the $\Lbz$-$\Tbz$ correlation (\citealt{2013ApJ...774..157D}b, for the calibration of 3D correlation, $L_{X}$-$T_{X}$-$L_{peak}$, see \citealt{2022MNRAS.514.1828D}a, \citealt{2023MNRAS.518.2201D}a, \citealt{2023ApJ...951...63D}b). Recently, low redshift SNe Ia data have been used to calibrate GRBs \citep{2008ApJ...685..354L, 2009MNRAS.400..775C, 2014ApJ...783..126P, 2021ApJ...920..135X}. In addition, an alternative method is to use Gaussian process with Observational Hubble Data Set (OHD) to calibrate GRB correlations \citep{2022ApJ...924...97W}.

Inspired by this point, we want to use the same method presented by \cite{2018ApJ...856....3Y} and \cite{2022ApJ...924...97W} to obtain the model-independed calibrated correlations. There are two main steps in the process: (1) use the Hubble parameter data to calibrate the luminosity distance ($d_L$) of low-redshift GRBs, and get the best fitting coefficients of empirical luminosity correlations with the calibrated low-redshift GRBs; (2) use the fitting results to calculate the model-independent distance modulus ($\mu_{\rm obs}$) of all selected GRBs, and then obtain the constraints on cosmological parameters.

The continuous function $h(x) = H(z)$ can be reconstructed using the Gaussian process (GP) method (see \cite{2018ApJ...856....3Y} for a more detailed explanation).
Using the $h(x)$, we can get the values of $H(z)$ for different redshifts, and the luminosity distance in this process can be expressed as
\begin{eqnarray}\label{eq:14}
\dl(z) = c(1+z)\int_{0}^{z} \frac{dz}{H(z)}.
\end{eqnarray}

In the calibration process, 36 $H(z)$ data covering the redshift range (0.07, 2.36) from \cite{2018ApJ...856....3Y} have been used. Using GP method, we can get the distances of GRBs in the redshift range of $z\lesssim 2.50$. A total of 23 GRBs in the sample satisfy this requirement. Based on equation (\ref{eq:14}), the calibrated luminosity distance can be used to recalculate the $\Lbz$ and $\Eiso$ of the low-redshift GRBs by using Equations (\ref{eq:4}) and (\ref{eq:7}). In this phase, we rederived calibrated empirical luminosity correlations using key parameters of 23 low-redshift GRBs. We then used the MCMC method to obtain the best fitting results. 

\subsection{Constraints on cosmological parameters with calibrated correlations }
\subsubsection{L-T correlation}
The MCMC method is used to get the best fitting results of 23 low-redshift GRBs as $k = -1.27\pm0.27$, $b = 2.20\pm0.30$ and $\sigma_{\rm int} = 0.71\pm0.11$.

For the flat ${\Lambda}$CDM model, the expression of distance modulus and luminosity distance is
\begin{equation}\label{eq:15}
\mu= 5\log\frac{\dl}{\rm{Mpc}} + 25 = 5\log\frac{\dl}{\rm{cm}} -97.45.
\end{equation}

At this stage, we have obtained the best fitting coefficients of calibrated Dainotti relation for all low-redshift GRBs. Next, we use the calibrated Dainotti relation to extrapolated to the $z\geq 2.50$ sample, called the high-redshift sample. In previous work, the Dainotti relation obtained by us can be expressed as $(\log{\Lbz} - 40) = b + k(\log{\Tbz} - 5)$.
By replacing $\dl$ with function $\sqrt{\Lbz/(4\pi F_{\rm b}(1+z)^{-4/3})}$ and combined with (\ref{eq:15}), we can obtain the observed distance modulus ($\mu_{\rm obs}$) of total 27 GRBs and the uncertainty ($\sigma_{\rm obs}$). The equation expression can be derived as follows:
\begin{equation}
\begin{aligned}
\mu_{\rm obs} &= \frac{5}{2} \left[{k(\log{\Tbz}-5)+b+40}\right. \\
&\left.{-\log{4\pi F_{\rm b}(1+z)^{-4/3}}}\right] -97.45,
\end{aligned}
\end{equation}
and
\begin{equation}
\begin{aligned}
\sigma_{\rm obs} &= \frac{5}{2} \left[{(\log{\Tbz} -5)^2 \sigma_{\rm k}^{2} + k^{2}\sigma_{\log{\Tbz}}^2+\sigma_{\rm b}^{2}} \right. \\
&\left.{+ (\frac{\sigma_{\rm F_b}}{F_{\rm b} \ln{10}})^{2} + \sigma_{\rm int}^{2}}\right]^{1/2}.
\end{aligned}
\end{equation}

The calculated results of $\mu_{\rm obs}$ are shown in Table 3. In order to obtain the constraints on cosmological parameters, one can write the likelihood function, $\mathcal{L}_{\rm{GRB}}(\theta)$, as follows \citep{2019MNRAS.486L..46A}
\begin{eqnarray}
		\mathcal{L}_{\rm{GRB}}(\theta)&=&\prod_{i=1}^{\mathcal{N}_{\rm{GRB}}}
    \frac{1}{\sqrt{2\pi}\sigma_{\rm obs}(z_{\rm i})} \exp
    \nonumber \\
   &\times&\left[-\frac{1}{2}\left( \frac{\mu_{\rm obs}(z_{\rm i})-\mu_{\rm{th}}(z_{\rm i},\theta)}{\sigma_{\rm obs}(z_{\rm i})}
    \right)^{2} \right],
\end{eqnarray}
where $\mu_{\rm{th}}(z_{\rm i},\theta)$ is the theoretical distance modulus calculated from equations (\ref{eq:3}) and (\ref{eq:15}) for a flat ${\Lambda}$CDM model.

Using the $\LT$ correlation and the model-independent $\mu_{\rm obs}$ and $\sigma_{\rm obs}$ of GRBs, we can obtain the constraints on cosmological parameters. For the flat ${\Lambda}$CDM model, we come to the conclusion $\om = 0.527\pm0.280$. For the non-flat ${\Lambda}$CDM model, the expression of the luminosity distance has changed, and it can be written as follows (\citealt{2007ApJ...667....1W}; \citealt{2013ApJ...774..157D}b):
\begin{equation}
d_{\rm L}=\left\{
\begin{array}{l}
\displaystyle
\frac{c(1+z)}{H_{0}\sqrt{-\Omega_{\rm k}}}\sin(\sqrt{-\Omega_{\rm k}}~ I), ~~~ \Omega_{\rm k}<0, \\
\displaystyle \frac{c(1+z)}{H_{\rm 0}}~I, ~~~~~~~~~~~~~~~~~~~~~ \Omega_{\rm k}=0,\\
\displaystyle
\frac{c(1+z)}{H_{0}\sqrt{\Omega_{\rm k}}}\sinh(\sqrt{\Omega_{\rm k}}~I),
~~~~~~\Omega_{\rm k}>0,\\
\end{array} \right.
\label{eqn:fc:}
\end{equation}
where
\begin{equation}
\Omega_{\rm k}=1-\om-\oL,
\end{equation}
and
\begin{equation}
I=\int_{0}^{z}\frac{dz}{\sqrt{(1+z)^{3}\om+\oL+(1+z)^{2}\Omega_{\rm k}}}.
\end{equation}

Perhaps due to the small number of our GRB samples, it does not have the ability to limit the non-flat ${\Lambda}$CDM model. In order to better constrain the cosmological models, combining GRB sample with more probes is a useful measure for us, such as Pantheon SNe Ia \citep{2018ApJ...859..101S} and CMB \citep{2016A&A...594A..13P,2020A&A...641A...6P}. The key to combining GRB samples with different probes is to effectively synthesize the likelihood functions of various probes.

The likelihood function of SNe Ia can be written as
\begin{eqnarray}
		\mathcal{L}_{\rm{SN}}(\theta)&=&\prod_{i=1}^{\mathcal{N}_{\rm{SN}}}
    \frac{1}{\sqrt{2\pi}\sigma_{\rm SN}(z_{\rm i})} \exp
    \nonumber \\
   &\times&\left[-\frac{1}{2}\left( \frac{\mu_{\rm SN}(z_{\rm i})-\mu_{\rm{th}}(z_{\rm i},\theta)}{\sigma_{\rm SN}(z_{\rm i})}
    \right)^{2} \right],
\end{eqnarray}
where $\mu_{\rm SN}(z_{\rm i})$ and $\sigma_{\rm SN}(z_{\rm i})$ are given by the Pantheon SNe Ia sample \citep{2018ApJ...859..101S}.

For the CMB, the shift parameter $R$ is considered. Under the circumstances, the likelihood function is
\begin{eqnarray}
		\mathcal{L}_{\rm{CMB}}(\theta)=\frac{1}{\sqrt{2\pi}\sigma_{\rm R}} \exp
   \left[-\frac{1}{2}\left( \frac{R(\theta)-R_{\rm{obs}}}{\sigma_{\rm R}}
    \right)^{2} \right],
\end{eqnarray}
where $R_{\rm obs}$ and $\sigma_{\rm R}$ can be acquired from the Planck-2015 data. Here we have set $R_{\rm obs}=1.7482\pm0.0048$ and $R_{\rm obs}=1.7474\pm0.0051$ for a flat and a non-flat universe, respectively \citep{2016PhRvD..94h3521W}.

In order to combine the GRB and the Pantheon samples to better limit cosmological parameters, we have used the joint likelihood function as $\mathcal{L}=\mathcal{L}_{\rm{GRB}}\mathcal{L}_{\rm{SN}}$. The optimal fitting results are $\om = 0.285\pm0.008$ for the flat ${\Lambda}$CDM model and $\om = 0.344\pm0.036$, $\oL = 0.788\pm0.037$ for the non-flat ${\Lambda}$CDM model. For including the CMB data, the likelihood function can be succinctly described as $\mathcal{L}=\mathcal{L}_{\rm{GRB}}\mathcal{L}_{\rm{SN}}\mathcal{L}_{\rm{CMB}}$. The constraint condition of flat ${\Lambda}$CDM model is $\om = 0.297\pm0.006$. In the case of non-flat ${\Lambda}$CDM model, the result is $\om = 0.283\pm0.008$, $\oL = 0.711\pm0.006$, shown in Figure 4(a). The constraints obtained by combining GRBs data with other probes data are better than those obtained by GRBs data alone. In Figure 4(b), the Hubble diagram constructed by using calibrated distance modulus ($\mu_{\rm obs}$) is exhibited. It can be notice that the error bars of the GRB distance modulus are still a bit too large, implying that $\LT$ correlation of GRBs alone cannot accurately limit cosmological parameters. The final results are summarized in Table 4.

\subsubsection{L-T-E correlation}

In this part, we explored the possibility of using the $\LTE$ correlation to constraint cosmological parameters. Using the low-redshift GRBs, the best fitting results of the related parameters in $\LTE$ correlation are $a = 1.84\pm0.36$, $b = -1.03\pm0.29$, $c = 0.18\pm0.13 $ and $\sigma_{\rm int} = 0.70\pm0.10$.

The model-independent $\mu_{\rm obs}$ and $\sigma_{\rm obs}$ can be written as
\begin{equation}
\begin{aligned}
	\mu_{\rm obs} &= \frac{5}{2(1-c)} \left[ {a +b(\log{\Tbz}-5)+c(\log\frac{4\pi S_{\rm bolo}}{1+z}-52)}\right. \\
 &\left.{ -\log{4\pi F_{\rm b}(1+z)^{-4/3}}+ 40 }\right] -97.45,
\end{aligned}
\end{equation}
and
\begin{eqnarray}
\lefteqn{ \sigma_{\rm obs} = \frac{5}{2(1-c)} \Bigg\{ \ext^2
+\sigma_{\rm a}^2 +\sigma_{\rm b}^2(\log{\Tbz}-5)^2 +b^{2}\sigma_{\log{\Tbz}}^2}
\nonumber\\
&+(\frac{\sigma_{\rm c}}{1-c})^2 \left[ a+b(\log{\Tbz}-5)-\log\frac{4\pi F_{\rm b}}{(1+z)^{4/3}}
+\log\frac{4\pi S_{\rm bolo}}{1 + z}-12\right]^{2}
\nonumber\\
&+(\frac{\sigma_{\rm F_{b}}}{\ F_{\rm b} \ln10})^2  \Bigg\}^{1/2}.
\end{eqnarray}

The calculation results are shown in Table 3. Finally, the best fitting result is $\om = 0.519\pm0.314$ for the flat ${\Lambda}$CDM model only by GRB sample. The constraints for including the SNe Ia and CMB data are $\om = 0.298\pm0.006$ for the flat ${\Lambda}$CDM model and $\om = 0.283\pm0.008$, $\oL = 0.711\pm0.007$ for the non-flat ${\Lambda}$CDM model. These are summarized in Table 5. The best fitting results and calibrated Hubble diagram are shown in Figure 5.

\subsubsection{L-T-$E_{\rm p,i}$ correlation}

For $\LTEpi$ correlation, we also obtained the best fitting results using 23 low-redshift GRBs, $a = 2.20\pm0.42$, $b = -1.25\pm0.31$, $c = -0.03\pm0.37$ and $\sigma_{\rm int} = 0.73\pm0.11$. The model-independent $\mu_{obs}$ and the uncertainty $\sigma_{\rm obs}$ are derived as
\begin{equation}
\begin{aligned}
	\mu_{\rm obs} &= \frac{5}{2} \left[ {a +b(\log{\Tbz}-5)+c(\log\Epi-2) + 40}\right. \\
 &\left.{ -\log{4\pi F_{\rm b}(1+z)^{-4/3}} }\right] -97.45,
\end{aligned}
\end{equation}
and
\begin{equation}
\begin{aligned}
	\sigma_{\rm obs}&=\frac{5}{2}\left[{\ext^2 +\sigma_{\rm a}^2+\sigma_{\rm b}^2(\log{\Tbz}-5)^2 +b^{2}\sigma_{\log{\Tbz}}^2 }\right. \\
 &\left.{+{\sigma_{\rm c}^2(\log{\Epi}-2)}^2+c^2{\sigma^2_{\log{\Epi}}} +(\frac{\sigma_{\rm F_{ b}}}{\ F_{\rm b} \ln10})^2 }\right]^{1/2}.
\end{aligned}
\end{equation}

The final results are shown in Table 3. The best fitting result by GRB data alone is $\om = 0.578\pm0.325$. Combining GRB sample with other probes, we obtained $\om = 0.298\pm0.005$ for the flat ${\Lambda}$CDM model and $\om = 0.283\pm0.008$, $\oL = 0.711\pm0.006$ for the non-flat ${\Lambda}$CDM model, respectively. The results are displayed in Table 6. The constraint results diagram and calibrated Hubble diagram are shown in Figure 6.

\subsection{Comparison of the results with the fundamental plane relations in optical and X-rays.}

In the case of GRBs, there have been many attempts to treat them as a standard candle. The X-ray plateaus 2D Dainotti relation was the first attempt to use afterglow correlation as a tool for studying cosmology. In addition, much of the works to limit cosmological parameters have been done in conjunction with the 3D Dainotti fundamental plane relation of GRBs. \cite{2022MNRAS.514.1828D}a standardized GRBs using 3D Dainotti relation of X-ray samples $L_{X}$-$T_{X}$-$L_{peak}$. By combining SNe Ia and GRBs, they obtained the  constraint of $\om = 0.299\pm0.009$ for a flat ${\Lambda}$CDM cosmology, and found that 3D optical Dainotti correlation can also be used to measure $\om$. In the paper of \cite{2022PASJ...74.1095D}b, 3D Dainotti correlation of the platinum
sample (including 50 long GRBs) corrected for selection biases and redshift evolution were used to constrain the cosmological parameters together with SNe Ia and BAOs. Their results extended the distance ladder to $z = 5$, indicating the importance of GRBs. A correlation study of \cite{2023MNRAS.518.2201D}a indicates that a lower central value for the intrinsic scatter
$\sigma_{\rm int} = 0.18\pm0.07$ can be obtained by using the corrected 3D Dainotti correlation of platinum sample. To improve the accuracy of constrained cosmology, \cite{2022MNRAS.512..439C}a combined 50 standardized platinum Dainotti-correlated GRBs with 101 selected standardized Amati-correlated GRBs to constrain the cosmological parameters. This set of samples probed the higher redshift range of $z\sim2.3-8.2$, but not precise enough. By using the updated data, two important cosmological parameters are constrained: the Hubble constant $H_{0} = 69.8\pm1.3$ km $\rm s^{-1}$ $\rm Mpc^{-1}$, and the non-relativistic matter density parameter $\om = 0.288\pm0.017$ \citep{2023PhRvD.107j3521C}. \cite{2023ApJ...951...63D}b and \cite{2023MNRAS.521.3909B} applied a combination of SNe Ia, BAOs, QSOs, and GRBs to constrain the cosmological parameters, where the selection bias and redshift evolution are taken into account  for the Risaliti-Lusso relation of QSOs and 3D Dainotti correlation of GRB X-ray plateaus emission.

\begin{table*}[h]\footnotesize %
  \caption{Results of the Linear Regression Analysis for Radio Plateaus}
  \setlength{\tabcolsep}{1.0mm}{
  \begin{center}
  \begin{tabular}{ccccccc}
  \hline
  \hline
  Correlations & Expressions & $\sigma_{\rm int}$\\
  \hline
  $\Lbz$($\Tbz$) ~~~&  $\log(\Lbz/10^{40}\rm{erg~s^{-1}})=(2.13\pm0.24)+(-1.20\pm0.24)\log(\Tbz/10^{5}\rm{s})$       & ~~~~~ $0.66\pm0.09$ \\
  \hline
  $\Lbz$($\Tbz$, $\Eiso$) ~~~ &  $\log(\Lbz/10^{40}\rm{erg~s^{-1}})=(1.90\pm0.26)+(-1.01\pm0.24)\log(\Tbz/10^{5}\rm{s})$  &~~~~~ $0.63\pm0.09$\\
  &    $ ~~~~~~~~~~~~~~~~~+(0.18\pm0.09)\log(\Eiso/10^{52}\rm{erg})$        &      \\
  $\Lbz$($\Tbz$, $\Epi$) ~~~  &   $\log(\Lbz/10^{40}\rm{erg~s^{-1}})=(2.10\pm0.34)+(-1.18\pm0.27)\log(\Tbz/10^{5}\rm{s})$        & ~~~~~ $0.67\pm0.09$   \\
  &  $ ~~~~~~~~~~~~~~~+(0.05\pm0.28)\log(\Epi/10^{2}\rm{keV})$      &     \\
  \hline
\end{tabular}
\end{center}}
\end{table*}

\begin{table*}[h]\footnotesize   %
  \caption{The fitting results for our radio samples }
  \setlength{\tabcolsep}{1.0mm}{
  \begin{center}
  \begin{tabular}{cccccccccc}
  \hline
  \hline
    GRB  & $z$ &    $\LT$   & $\LTEiso$  &   $\LTEpi$ \\
       & & $\mu_{obs}$ &$\mu_{obs}$ &$\mu_{obs}$\\
  \hline
  171205A	&	0.0368	&	36.59$\pm$1.96	&	34.94$\pm$2.54	&	36.59$\pm$2.15	\\
  031203	&	0.105	&	41.42$\pm$2.44	&	40.98$\pm$2.98	&	41.47$\pm$2.66	\\
  030329	&	0.168	&	36.91$\pm$2.01	&	36.31$\pm$2.47	&	36.93$\pm$2.21	\\
  020903	&	0.25	&	41.09$\pm$2.51	&	39.91$\pm$3.38	&	41.26$\pm$3.06	\\
  171010A	&	0.3285	&	42.96$\pm$1.94	&	43.44$\pm$2.57	&	42.94$\pm$2.16	\\
  160623A	&	0.367	&	39.26$\pm$3.08	&	38.53$\pm$3.45	&	39.21$\pm$3.27	\\
  190114C	&	0.425	&	43.24$\pm$1.97	&	43.80$\pm$2.59	&	43.19$\pm$2.34	\\
  211106A	&	0.5	&	43.91$\pm$2.19	&	43.72$\pm$2.71	&	43.93$\pm$2.49	\\
  050713B	&	0.55	&	44.10$\pm$2.08	&	44.43$\pm$2.61	&	44.12$\pm$2.30	\\
  070612A	&	0.617	&	40.65$\pm$2.27	&	40.76$\pm$2.82	&	40.69$\pm$2.53	\\
  991208	&	0.706	&	43.05$\pm$2.32	&	43.67$\pm$2.88	&	43.06$\pm$2.56	\\
  980703	&	0.966	&	44.18$\pm$2.03	&	44.33$\pm$2.54	&	44.15$\pm$2.31	\\
  071003	&	1.1	&	45.80$\pm$2.07	&	46.19$\pm$2.64	&	45.73$\pm$2.55	\\
  000418	&	1.119	&	43.65$\pm$2.04	&	43.76$\pm$2.53	&	43.65$\pm$2.28	\\
  191221B	&	1.148	&	42.29$\pm$1.92	&	41.77$\pm$2.37	&	42.22$\pm$2.27	\\
  141121A	&	1.47	&	45.57$\pm$2.06	&	45.84$\pm$2.57	&	45.58$\pm$2.27	\\
  010222	&	1.477	&	45.61$\pm$2.47	&	46.85$\pm$3.19	&	45.60$\pm$2.78	\\
  070125	&	1.548	&	44.00$\pm$2.10	&	44.52$\pm$2.66	&	43.97$\pm$2.47	\\
  990510	&	1.619	&	48.20$\pm$1.94	&	48.55$\pm$2.58	&	48.16$\pm$2.18	\\
  120326A	&	1.798	&	44.19$\pm$2.28	&	44.28$\pm$2.75	&	44.22$\pm$2.47	\\
  000926	&	2.039	&	47.19$\pm$1.96	&	47.84$\pm$2.63	&	47.17$\pm$2.20	\\
  111215A	&	2.06	&	43.63$\pm$1.96	&	43.14$\pm$2.40	&	43.61$\pm$2.19	\\
  021004	&	2.33	&	44.82$\pm$2.03	&	44.36$\pm$2.49	&	44.82$\pm$2.26	\\
  011030	&	3	&	47.26$\pm$2.00	&	47.29$\pm$2.48	&	47.31$\pm$2.23	\\
  090323	&	3.57	&	47.45$\pm$2.03	&	48.23$\pm$2.72	&	47.38$\pm$2.51	\\
  980329	&	3.9	&	46.36$\pm$2.04	&	47.24$\pm$2.72	&	46.33$\pm$2.43	\\
  140304A	&	5.283	&	49.59$\pm$1.94	&	49.55$\pm$2.48	&	49.52$\pm$2.29	\\
 \hline
\end{tabular}
\end{center}}
\end{table*}

\begin{table*}[h]\footnotesize %
  \caption{Constraints on cosmological parameters with $\LT$ correlation}
  \setlength{\tabcolsep}{1.0mm}{
  \begin{center}
  \begin{tabular}{ccccccc}
  \hline
  \hline
  Flat ${\Lambda}$CDM & $\om$\\
  \hline
  GRB &   $0.527\pm0.280$  \\
  GRB+SN &    $0.285\pm0.008$   \\
  GRB+SN+CMB  &   $0.297\pm0.006$   \\
  \hline
  Non-flat ${\Lambda}$CDM & $\om$  & $\oL$  \\
  \hline
  GRB+SN & $0.344\pm0.036$   &  $0.788\pm0.037$ \\
  GRB+SN+CMB  &  $0.283\pm0.008$  &  $0.711\pm0.006$  \\
  \hline
\end{tabular}
\end{center}}
\end{table*}

\begin{table*}[h]\footnotesize %
  \caption{Constraints on cosmological parameters with $\LTEiso$ correlation}
  \setlength{\tabcolsep}{1.0mm}{
  \begin{center}
  \begin{tabular}{ccccccc}
  \hline
  \hline
  Flat ${\Lambda}$CDM & $\om$\\
  \hline
  GRB &   $0.519\pm0.314$  \\
  GRB+SN &    $0.285\pm0.008$   \\
  GRB+SN+CMB  &   $0.298\pm0.006$     \\
  \hline
  Non-flat ${\Lambda}$CDM & $\om$  & $\oL$  \\
  \hline
  GRB+SN & $0.343\pm0.035$   &  $0.787\pm0.044$ \\
  GRB+SN+CMB  &  $0.283\pm0.008$  &   $0.711\pm0.007$ \\
  \hline
\end{tabular}
\end{center}}
\end{table*}

\begin{table*}[h]\footnotesize %
  \caption{Constraints on cosmological parameters with $\LTEpi$ correlation}
  \setlength{\tabcolsep}{1.0mm}{
  \begin{center}
  \begin{tabular}{ccccccc}
  \hline
  \hline
  Flat ${\Lambda}$CDM & $\om$\\
  \hline
  GRB &   $0.587\pm0.325$  \\
  GRB+SN &    $0.285\pm0.008$   \\
  GRB+SN+CMB  &   $0.298\pm0.005$     \\
  \hline
  Non-flat ${\Lambda}$CDM & $\om$  & $\oL$  \\
  \hline
  GRB+SN & $0.343\pm0.035$   &  $0.787\pm0.043$ \\
  GRB+SN+CMB  &    $0.283\pm0.008$  &   $0.711\pm0.006$    \\
  \hline
\end{tabular}
\end{center}}
\end{table*}

\begin{figure*}
	\centering
	\subfigure[GRB+SN+CMB constraints for non-flat ${\Lambda}$CDM.]{
		\includegraphics[width=0.38\textwidth,angle=0]{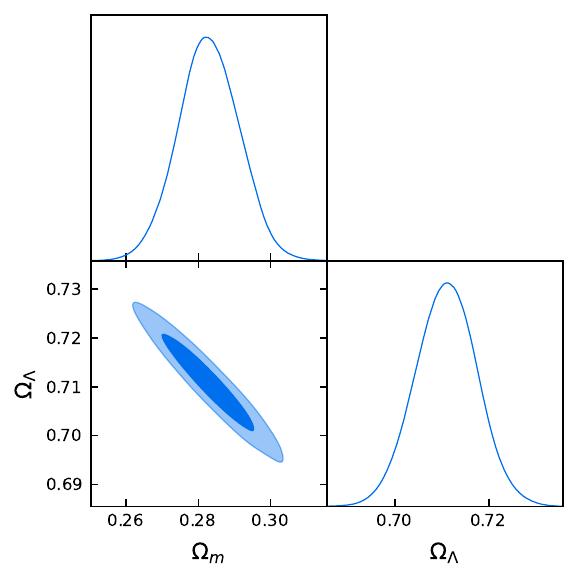}}
	\hfill
	\centering
	\subfigure[The Hubble diagram of GRBs and supernovae.]{
		\includegraphics[width=0.45\textwidth,angle=0]{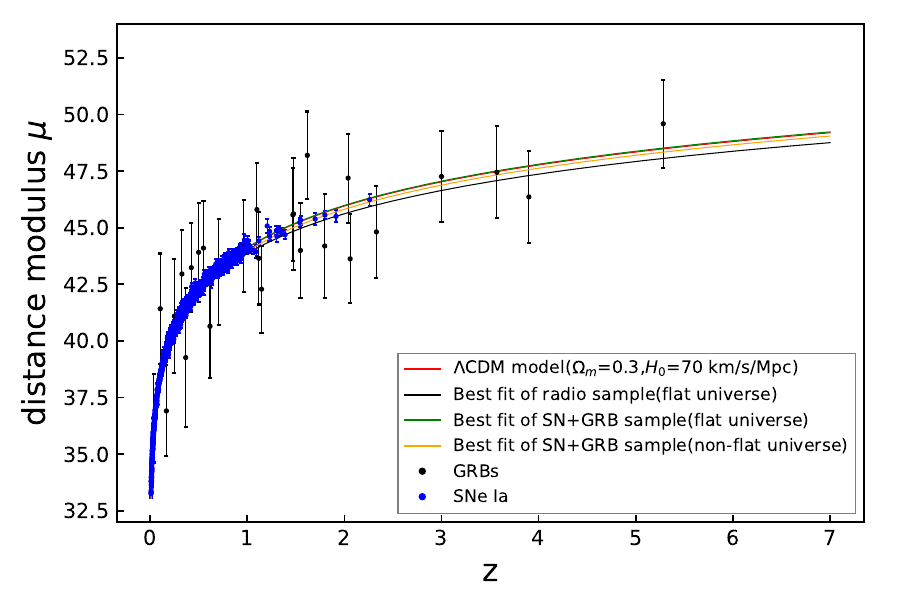}}
	\caption{Constraints on cosmological parameters and calibrated GRB Hubble diagram for $\LT$ correlation. (a) corner plots for the constraints for the joint radio GRB, SNe Ia and CMB samples. The confidence regions of the parameters $\Omega_{\rm m}$ and $\Omega_{\rm \Lambda}$ are 1 $\sigma$ and 2 $\sigma$ from the inner to the outer; (b) Blue points are supernovae from the Pantheon sample. Black points are 27 GRBs with plateau phase. The red solid line is a flat $\Lambda$CDM model with $\om = 0.3$ and $H_0$ = 70.0 km s$^{-1}$ Mpc$^{-1}$. For flat $\Lambda$CDM model, the black and green solid line are the best fit of calibrated GRB sample and SN+GRB sample, respectively. The best fit from SN+GRB sample for non-flat $\Lambda$CDM model is shown as orange line.}
\end{figure*}

\begin{figure*}
	\centering
	\subfigure[GRB+SN+CMB constraints for non-flat ${\Lambda}$CDM.]{
		\includegraphics[width=0.38\textwidth,angle=0]{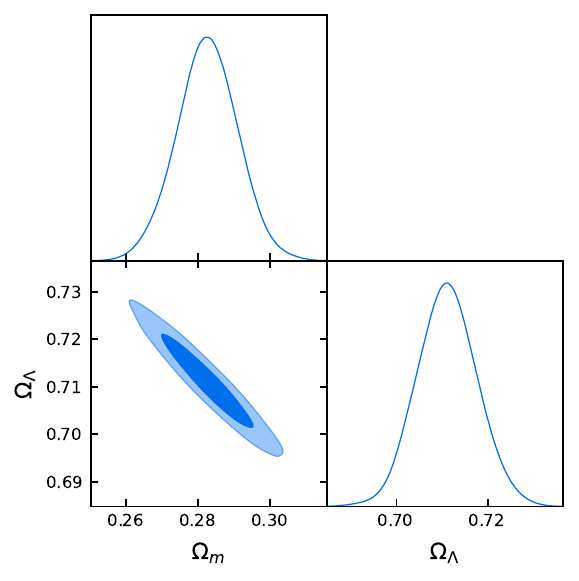}}
	\hfill
	\centering
	\subfigure[The Hubble diagram of GRBs and supernovae.]{
		\includegraphics[width=0.45\textwidth,angle=0]{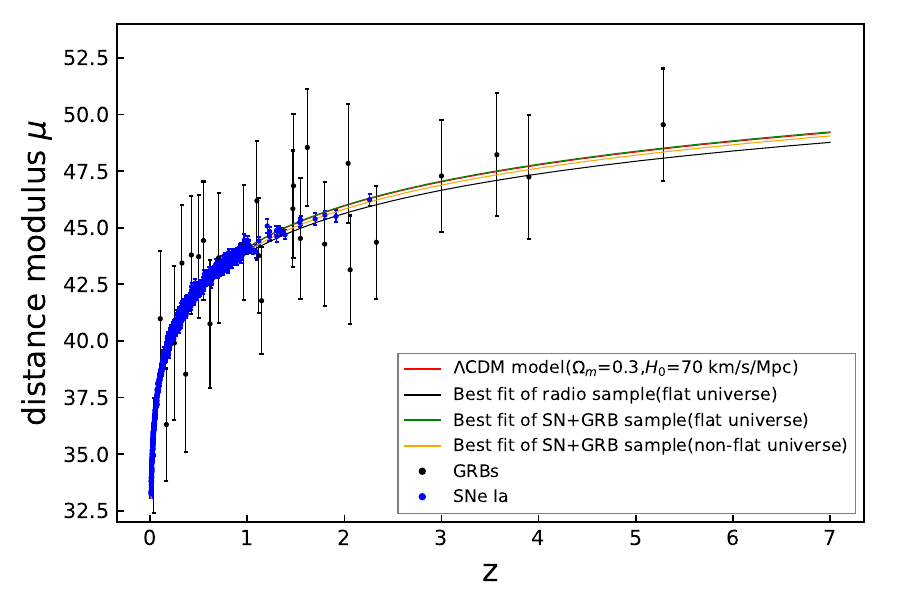}}
	\caption{Constraints on cosmological parameters and calibrated GRB Hubble diagram for $\LTEiso$ correlation.}	
\end{figure*}

\begin{figure*}
	\centering
	\subfigure[GRB+SN+CMB constraints for non-flat ${\Lambda}$CDM.]{
		\includegraphics[width=0.38\textwidth,angle=0]{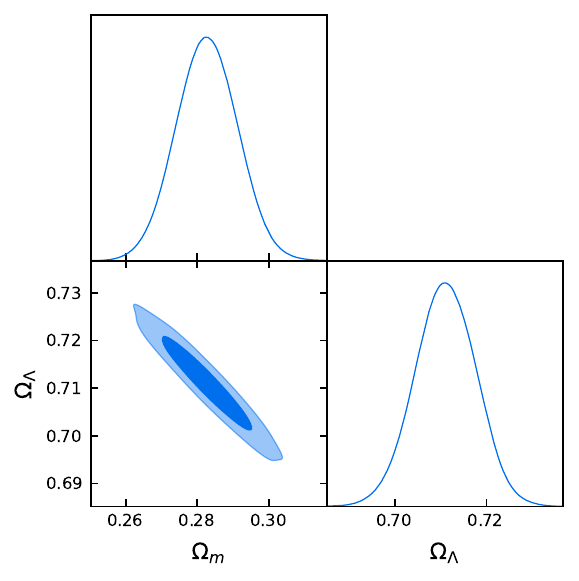}}
	\hfill
	\centering
	\subfigure[The Hubble diagram of GRBs and supernovae.]{
		\includegraphics[width=0.45\textwidth,angle=0]{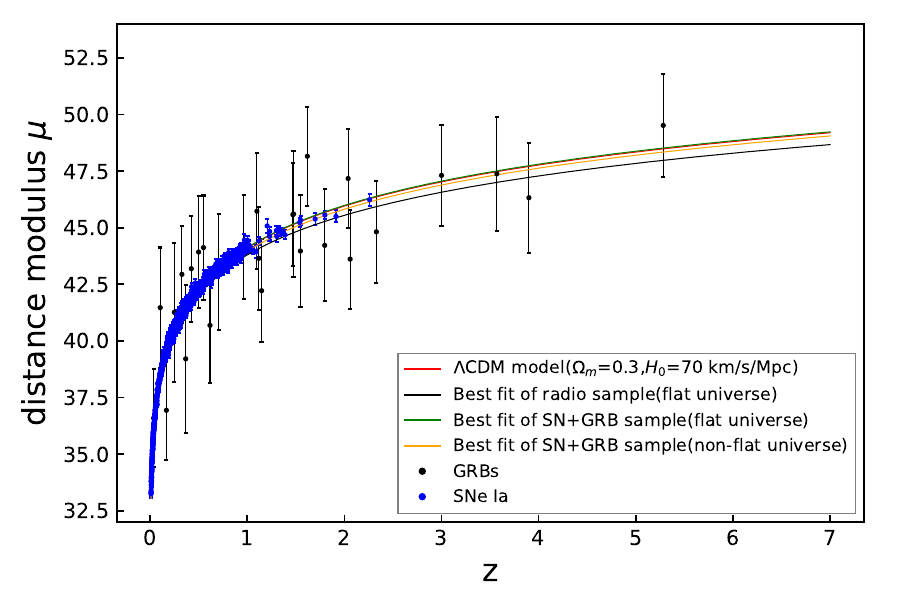}}
	\caption{Constraints on cosmological parameters and calibrated GRB Hubble diagram for $\LTEpi$ correlation.}	
\end{figure*}

\section{Discussion and Conclusion}

In order to investigate the properties of radio plateau afterglows, we systematically selected 27 GRBs with radio plateau and fitted them by a empirical smooth broken power-law function. We obtained the radio plateau parameters such as the break time ($\Tbz$), the radio flux of the break time ($F_b$), and the temporal indexes ($\alpha_1$ and $\alpha_2$). The criteria of radio plateau sample need to be satisfy $0 < |\alpha_{\rm 1}| < 0.5$. The break time of radio plateaus is generally later than that of X-ray and optical plateaus, and the corresponding break luminosity ($\Lbz$) is several orders of magnitude lower than that of X-ray and optical afterglow plateaus. We conducted a statistical analysis based on the selected samples. Firstly, we explored the Dainotti relation and found that the radio plateaus present a relatively compact negative correlation, $\Lbz \propto \Tbz^{-1.20\pm0.24}$. Secondly, we added the isotropic energy $\Eiso$ and the peak energy $\Epi$ into the correlation of $\Lbz$-$\Tbz$ for radio plateaus, and paid attention to that GRBs with an obvious plateau phase in radio bands also existed the new three-parameter correlations, $\Lbz \propto \Tbz^{-1.01 \pm 0.24} \Eiso^{0.18 \pm 0.09}$ and $\Lbz \propto \Tbz^{-1.18 \pm 0.27} \Epi^{0.05 \pm 0.28}$.

Based on the works of predecessors, several different models have been proposed to explain the plateau. It is thought to be generated by the energy injection of a rapidly rotating millisecond magnetar locating in the heart of the GRB remnant\citep{2001ApJ...552L..35Z,2004ApJ...606.1000D,2013MNRAS.430.1061R,2014ApJ...785L...6S,2015ApJ...813...92R,
2018MNRAS.480.4402L}. According to this model, the millisecond magnetar spins down and loses rotational energy after the prompt emission. Meanwhile, a Poynting flux or electron-positron wind is generated and energy is injected into the the external shock. Our results are less consistent with that of X-ray and optical afterglows, suggesting that the radio plateau may have a different origin.

GRB 141121A and GRB 070125 are two special cases in our sample. GRB 141121A not only has a plateau phase in the radio band, but also shows a plateau phase in the X-ray band. Interestingly, GRB 070125 has a plateau phase in both optical and radio bands, implying that central engine energy injection may be responsible for the plateau phase characteristics of this type of GRBs, and central engine powered energy injection should be presented in all bands. However, due to the difference of radiation efficiency, energy injection may be mostly manifested in the plateau phase of the X-ray band with high radiation efficiency, followed by the optical band. The detection probability of the radio plateaus is much smaller than that of X-ray and optical plateaus. Upon accomplishing this article, we were drawn attention to \cite{2023MNRAS.519.4670L}, who investigated the theoretical interpretation of radio afterglow under different frameworks, namely the standard fireball model and the energy injection model. Their research showed that considering the presence of energy injection does not necessarily make the radio data better compatibility with standard fireball model, but gamma-ray bursts with radio plateau seem more possibly to be consistent with the standard fireball model.

Two three-parameter empirical luminosity correlations of the radio plateau data are also exist, suggesting that the central engine of the burst should be related to the surrounding environment, or the internal radiation mechanism plays a predominant role in the correlations. For the former speculation, the deceleration of the GRB jets may be responsible for the breaks. The radio plateau phase appears relatively late, when the environment medium is more complex. Radio plateaus may result from the interaction of jets with more complex ambient media. For the latter speculation, the typical frequencies crossing the observational band may be a reasonable hypothesis that causes the breaks of the radio afterglows.

We also used standardised GRBs as standard candle to limit cosmological parameters with empirical luminosity correlations in the radio bands. The method of correlation calibration in this paper, known as calibration on SNe Ia, is also employed by \cite{2014ApJ...783..126P} and \cite{2023MNRAS.518.2201D}a. Our results show that constraint results are better for the flat $\Lambda$CDM model. For $\LT$ correlation, the constraint result is $\om = 0.527\pm0.280$ for the flat ${\Lambda}$CDM model using only GRB data. For the data of GRB+SN+CMB, the constraints are $\om = 0.297\pm0.006$ for the flat ${\Lambda}$CDM model, and $\om = 0.283\pm0.008$, $\oL = 0.711\pm0.006$ for the non-flat ${\Lambda}$CDM model.  From all the Hubble diagram, we found that the error bars of the GRB distance modulus are still too large. This may be due to the influence of the large degree of intrinsic scatter of the correlation. It should also be mentioned that the empirical luminosity correlations derived here is still not a very compact relation, due to the relatively large intrinsic scatter of $\sigma_{\rm int} \sim 0.6$. This suggests that the selected GRB sample alone currently cannot accurately limit cosmological parameters.

\section{Acknowledgments}
We thank the referee for very helpful suggestion and comments. This work is supported by the National Natural Science Foundation of China (Grant Nos. U2038106 and 12273009), the Shandong Provincial Natural Science Foundation (Grant No. ZR2021MA021), Jiangsu Funding Program for Excellent Postdoctoral Talent (20220ZB59) and Project funded by China Postdoctoral Science Foundation (2022M721561).

{}
\end{document}